\documentclass[11pt,twoside,letterpaper]{article} 
   \usepackage{amsfonts}
   \usepackage[psamsfonts]{amssymb}
   \usepackage{amsmath}
\usepackage{secdot}
   \usepackage[english]{babel}
\usepackage{titlesec}
\usepackage{times,fancyhdr}

\usepackage{amsfonts}
\usepackage{amssymb}
\usepackage{graphicx}
\usepackage{amsmath}%
\def\tablename{Table}
\makeatletter
\renewcommand{\fnum@table}[1]{\bfseries\tablename~\thetable.}
\makeatother
\def\figurename{Figure}
\makeatletter
\renewcommand{\fnum@figure}[1]{\figurename~\thefigure.}
\makeatother

\begin{document}

\title{
{\begin{flushleft}
\vskip 0.45in
{\normalsize\bfseries\textit{Chapter~ }}
\end{flushleft}
\vskip 0.45in \bfseries\scshape  Newtonian Limit of Einsteinian Gravity: From Dynamics of Solar System to Dynamics of Stars in Spiral Galaxies}}
\author{\bfseries\itshape  Arkady L. Kholodenko\thanks{Email address: string@clemson.edu}\\
375 H.L.Hunter Laboratories,\\
Clemson\ University, Clemson, SC 29634-0973, USA }

 \date{}
\maketitle 
\thispagestyle{empty} 
\thispagestyle{fancy} 
\fancyhead{}
\fancyhead[L]{In:   \\ 
Editor:  } 
\fancyhead[R]{ISBN:   \\
\copyright Science Publishers, Inc.} \fancyfoot{}
\renewcommand{\headrulewidth}{0pt}

\begin{abstract}
Attempts to merge Einsteinian gravity with Newonian run into
logical inconsistencies coming from the fact that in Newton's gravity
time is absolute and \ the speed of propagation of gravity is assumed to be
infinite. Such an assumption was in a focus of attention of many scientists
in 19th century interested in finding out if, indeed, the speed of propagation
of gravity is infinite. For this purpose, by analogy with electrodynamics,
some retarded potentials replacing Newtonian were suggested. Using one
of such potentials Gerber correctly calculated the perihelion shift for
Mercury in 1902. However, subsequent attempts at calculation of
bending of light using Gerber-style calculations were not successful.
Recently Gin\'{e} (Chaos, Solitons and Fractals \textbf{42, }1893\textbf{
(}2009\textbf{)) }was able 
to reobtain both the perihelion shift and the bending of light using retarded
potential. His equations \ however are not those obtained by Einstein and
his results coincide with those by Einstein only at the\ level of leading
order terms of infinite series expansions. The obtained differential
equations of motion are of delay-type. When applied to two-body
dynamics such equations lead to orbital quantization. In this work,
Einsteinian approach is being used to reproduce this quantization.
Numerous arguments justifying the superiority of Einsteinian approach,
including uses of the Bertrand spacetimes for description of motion of
stars around the galaxy center are provided. Bertrand spacetimes are being
studied currently in connection with issues related to dark matter existence.
The developed formalism is tested \ by calculating the number of allowed
stable orbits for planets and \ those for regular satellites of heavy planets
resulting in reasonable\ agreement with \ observational data. The paper also
discusses possible quantum mechanical nature of \ rings of heavy planets
as well as of rotation curves of stars in spiral galaxies.
\end{abstract}

\section{Introduction}

\subsection{General Remarks}

After almost hundred years since general relativity was formulated in its
final form the Newtonian limit of Einsteinian relativity is still a hot topic
of research \ e.g. see [$1,2$] and references therein. The problem stems from
attempts to reconcile irreconcilable. For instance, in Einsteinian universe
speed of propagation \ of gravity is finite and is assumed to be that for the
speed of light while in Newtonian it is infinite. Furthermore, as we
demonstrated in [$3-5$], at least formally, in two dimensions the Einteinian
gravity is well defined/behaved while the Newtonian is not. \ These facts
bring \ to life a whole array of recipes of either to modify Einsteinian or
Newtonian gravity or both, e.g. [$2,6$]. The history of making such a recipes
can be traced all the way back to 1804 when Soldner calculated bending of
light using Newtonian mechanics [$7$] and got the value for the angle of
deflection which is half of that obtained more than century later by Einstein.
According to [$8$] calculation of (some or all)\ orbital precession for the
Mercury had been ``a fairly popular activity in 1890's for physicists''. Among
these, the most notable is the calculation by Paul Gerber published in 1902
and reprinted in Annalen der Physik in 1917\footnote{E.g. see note on Paul
Gerber in Wikipedia\newline
 http://en.wikipedia.org/wiki/Paul\_Gerber}. Since Gerber's paper came out long
before Einstein's with correct result for the perihelion shift for Mercury,
Einstein was apparently aware of his result. Republication of Gerber's result
in Annalen der Physik in 1917 was meant to demonstrate Einstein's plagiarism.
The situation in this case is very similar to that in atomic mechanics when
Rutherford measured the \ differential cross-section \ for scattering of
electrons from Hydrogen atom. The calculated\ by means of classical mechanics
\ value of the cross-section \ nicely matched experimental data. Everybody
knows what happened next. In the case of relativity, Gerber was trying to use
the velocity -dependent potential, whose origins remained obscure, to account
for finite speed of propagation of gravity. Actually, his aim was not only to
calculate the shift correctly but to use this calculation in order to figure
out the speed of gravity. In 2010 this issue was still under active
investigation [$9$]. Use of Gerber's potential \ in calculations of bending of
light not only produces result which is twice that known experimentally and
correctly reproduced by Einstein [$8$] but, even more importantly, raises many
questions regarding the way calculations were made. Specifically, in
Einstein's calculations the value for angular momentum for such a calculation
is infinite while in Newtonian-type calculations [$7,8$] it is obviously
finite. It is possible to fiddle with Gerber's equations and to use
(artificially) the infinite value for the angular momentum. In which case one
obtains the value for bending which is 1.5 bigger than that obtained by by
Einstein. Using finite \ angular momentum produces already mentioned result
which is twice bigger.
 
\pagestyle{fancy} 
\fancyhead{} 
\fancyhead[EC]{Arkady L. Kholodenko} 
\fancyhead[EL,OR]{\thepage}
\fancyhead[OC]{Newtonian Limit of Einsteinian Gravity and Dynamics of Solar System}      
\fancyfoot{}
\renewcommand\headrulewidth{0.5pt}

Photons are quantum objects and their treatment in general relativity is
markedly different from that in Newton's mechanics. The bending of light
obtained in both Newtonian and Einsteinian theories (perhaps with some non
removable \ numerical discrepancies) is anticipated (recall that Soldner
published his results for bending in 1804). Much more striking is not bending
of light but existence of closed circular orbits for light predicted by
general relativity. By rotating such an orbit around its symmetry axis one
obtains what is known as ``photon sphere'' [$10$]. \ Even though such a sphere
should exist around every black hole, it is difficult \ apparently to probe it
experimentally. This difficulty is eliminated recently when it became possible
to model practically all known effects of general relativity in the laboratory
[$11-13$]. Not only this is of intrinsic interest scientifically, but also
this modelling has became a thriving area for designing of all kinds of
\ photo and other devices in which the effects of general relativity are to be
used commercially. In this work we would like to discuss those aspects of
general relativity which are related to celestial mechanics, not just to
calculations of perihelion shift for Mercury. Incidentally, the results of
such studies can also be tested both in the sky and in the lab. \ In
particular, since the photon sphere can be reproduced in the lab [$11$], it
can be analyzed both semiclassically (at the level of geometric optics) and
quantum mechanically. In the last case one is confronted with the following
problem: Can it be that only the photon sphere orbits could be treated quantum
mechanically while the rest of \ massive orbits (geodesics) strictly
classically? \ In this paper we argue in favor of treating both cases quantum
mechanically. \ 

Results and methods of this work differ substantially \ from that by
Gin\'{e}$^{{}}.$ In a series of papers culminating in [2] he attempted to
improve Gerber's calculations in order to reproduce Einstein's results for
both the perihelion shift for Mercury and for bending of light. While formally
succeeding in this task, he run across the delay differential equations
replacing more familiar Newton's \ ordinary differential equations valid
\ only in the absolute time. \ Methods of obtaining solutions of these delay
equations are nontrivial and lead quite naturally to quantization. He checked
his calculations using classical mechanics model of hydrogen atom accounting
for delay(s) and got known quantum mechanical spectrum for hydrogen bound
states. Next, he applied the same methods to the gravitational analog of
hydrogen atom and got spectrum reproducing the Titius-Bode law of planetary
distances. This law is also going to be discussed in this work in Section V.
Gin\'{e}$^{{}}$ noticed, that in the case of gravitational quantization the
Planck constant should be replaced by its celestial mechanics
analog/equivalent. In this work we shall reach the same conclusion. Given
this, his works are not without flaws. First, his equations of motion used for
calculation of perihelion shift and for bending of light are not those used by
Einstein for the same tasks. The obtained results agree with those by Einstein
only in the leading terms. Second, the Titius-Bode law obtained in his work
[14], and also reproduced by many authors cited in this paper is flawed. It is
working well only for the planets closest to the Sun and is becoming
increasingly inaccurate for more distant planets. More important is the fact
that both Gin\'{e} and the rest of authors do not account for the fact that
the number of planets around the Sun is finite. The same is true for the
number of satellites of heavy planets. Unlike the case of Hydrogen atom, where
the electroneutrality forbids more than one electron to be around proton, so
that all excited states are either empty or can be occupied by the same
electron excited from the ground state, in the sky there is no
electroneutrality. So,  if one believes in the validity of the Titius-Bode
law, one should anticipate a countable infinity of planets on \ a countable
infinity of allowed orbits. This is not observed and, therefore, is physically
meaningless. Notice also, that classically the photon sphere can also
accumulate unrestricted number of photons so that the observed mass of the
black hole should grow in observer's time without bound. This is also
physically meaningless. The photon devices made in the laboratory should
provide an upper bound on the density of photons which photon sphere is
capable of accommodating.

The ``quantization process'' applied to celestial mechanics had been initiated
not only in calculations by Gin\'{e} and other authors (mentioned in his
paper), it had also began in recent papers [$15-18$] discussing problems of
space travel within Solar System. Mathematically, these problems are analogous
to those encountered in quantum treatment \ of chemical kinetics of polyatomic
molecules. The following quotation from the paper by Porter and Cvitanovi\v{c}
[15] nicely illustrates the essence of these problems. `` Almost perfect
parallel between the governing equations of atomic physics and celestial
mechanics implies that the transport mechanism for these two situations is
virtually identical: on the celestial scale, transport takes a spacecraft from
one Lagrange point\footnote{That is point of equilibrium.} to another until it
reaches its desired destination\footnote{E.g. see also paper by Convay at al
\ [17] for details.}. On the atomic scale, the same type of trajectory
transports an electron initially trapped near the atom across the escape
threshold (in chemical parlance, across a ``transition state''), never to
return. The orbits used to design space missions thus also determine the
ionization rates of atoms and chemical reaction rates of molecules''. This
statement is nicely illustrated in the paper by Jaffe et al [18] in which it
is reported that the \textsl{transition state theory developed initially in
chemistry} (to describe the rates of chemical reactions) \textsl{is}
\textsl{working actually better in celestial mechanics} where the discrepancy
between the chemical theory and numerical simulations (done for celestial
mechanics transport problems) is less than 1\%. The current status of
transition state theory at the quantum and classical levels in chemistry is
nicely described in the recent book by Micha and Burghardt [$19$].

Before discussing the organization of the rest of this paper we would like to
mention the following empirical facts. If $M_{\odot}$ is the mass of the Sun
(or, respectively, heavy planet such as Jupiter, Saturn, etc.) and $m_{i}$ be
the mass of an i-th planet (respectively, an i-th satellite of heavy
planet)\footnote{For planets $i=1\div9$ while for satellites of heavy planets
this number is different as explained below, in the text.}, make the ratio
$r_{i}=\frac{m_{i}}{m_{i}+M_{\odot}}$ . The analogous ratios can be
constructed for respective heavy planets (Jupiter, Saturn, Uranus, Neptune)
and for any of their satellites. The observational data indicate that with
only two exceptions: Earth-Moon (for which $r\sim10^{-2}),$ and Pluto-Charon
(for which $r\sim10^{-1}),$ all other ratios in the Solar System are of order
$10^{-6}-10^{-3}$[20\textbf{]}. Under such circumstances the center of mass of
such a binary system practically coincides with that for $M_{\odot}$. And if
this is so, then the respective trajectories can be treated as geodesics.
Hence, not only motion of the Mercury can be treated in\ this way, as it was
done by Einstein, but also motion of almost any
satellite\footnote{Regrettably, not our Moon! \ \ The description of dynamics
of Moon is similar to that for rings of heavy planets (to be discussed in
Section V) and, as such, is also quantizable.} in the Solar System! These
empirical facts, plus those discussed in Section V, are compelling enough to
warrant calculations accounting for \ the superiority of Einsteinian gravity
over Newtonian already at the scales of our Solar System. As stated already in
$[3-5$], there are no well defined 2 dimensional Newtonian gravity but the
Einsteinian gravity is alive and well in 2 dimensions! That is, it is
mathematically well defined. This fact cannot be by-passed by fixing Newtonian
mechanics with help of time-dependent potentials.\bigskip

\subsection{Organization of the Rest of the Paper}

The paper is made of six sections and three appendices. In Section 2 we
provide some historical discussion, beginning with works of Laplace and
Poincare$^{\prime}$ on celestial mechanics. When these results are
reinterpreted in modern terms \ and superimposed with the latest combinatorial
formulations of quantum mechanics, they provide a foundation for thinking
about \ dynamics of Solar system quantum mechanically. These general results
are based on classical mechanics formalism. Superficially, such a formalism is
unable to treat photons (or neutrinos, etc.) and particles on the same footing
systematically (with the exception of the geometrical optics limit). Only
after discovery of general relativity this had become possible. Thus, in
Section 3 classical results of Section 2 are reanalyzed using formalism of
general relativity. In this section we provide detailed arguments in favor of
the quantum nature of the motion of planets on geodesics. Such a conclusion is
compatible with results of Section 2. Section 3 can be looked upon as an
analog of the mathematical proof of existence. To convert these abstract
results into numbers \ requires development of this formalism. It is presented
in Sections 4 and 5. In Section 4 we argue that the formalism of quantum
mechanics in its conventional form is not useful for development of quantum
celestial mechanics. In the same section we extend this formalism in order to
make it compatible with transformations normally used in general relativity.
Based on formalism developed in Section 4, we present the actual numerical
calculations in Section 5. Before doing so we briefly discuss the Bertrand
spaces. Such spaces were recently sucsessfully used for computation of the
rotation curves for stars in the spiral galaxies. Observatins show that
typical stars (including Sun in our galaxy) rotate around the galactic
center on circular orbits. The fact \ that the orbits are circular suggests
that these orbits might be of quantum origin. In section 5 we begin our
quantization program by developing the formalism both for planets and for
satellites of heavy planets. Our efforts culminate in Table 2. In it we
compare our \ analytical calculations of the available number of \ stable orbits
for planets and of\ the number of stable orbits for regular\footnote{The word
``regular'' is defined in Section 5} satellites of heavy planets with the
empirically available information. \ The obtained theoretical results are in
\ strikingly good agreement with the empirically observed data. They are in
accord with the quantum mechanical rules for filling the \ stable orbits
formulated in Section 5. Having these results \ obtained, we go on \ with
quantization. In the same section we present calculations demonstrating
quantum nature of rings around all heavy planets. Our treatment is compatible
with the requirements of general relativity. Finally, using methods of contact
geometry in addition to the already developed\ quantum mechanical formalism, we
obtain analytically the expression for the rotation curve for\ representative
stars rotating around the galactic center. The obtained theoretical result is in
remarkable agreement with the experimentaly obtained. It further supports the
existence of dark matter in our Universe. Section 6 is devoted to \ a brief
discussion  \bigskip

\section{Role of General Relativity and Quantum Mechanics in Solar System Dynamics (General Discussion)}

\subsection{From Laplace to Einstein  via Poincare${^{\prime}}$}

Even though classical Hamiltonians for Coulombic and Newtonian potentials look
\textsl{almost} the same, they are far from being \textsl{exactly} the same.
In the classical Hamiltonians for multielectron atoms \ all electron masses
are the same, while for the Solar system the masses of all satellites are
different. \ \ Using general relativity such difference can be made non
existent, since all planets/satellites are moving on geodesics, that is they
do not interact with each other. It is just the empirical fact coming from a
huge disparity in masses of the Sun and the planets (or the heavy planet and
its satellites) noticed already in section 1. In view of such mass disparity
one can proceed with formal quantization of both systems using the same
formalism. To explain how this happens, we begin with two- body Kepler problem
treated in representative physics textbooks [$21$]. Such treatments tend to
ignore the equivalence principle- essential for the gravitational Kepler
problem and nonexistent for the Coulomb-type problems. Specifically, the
description of general relativity in Vol.2 of the \ world-famous
Landau-Lifshitz course in theoretical physics [$22$] begins with the
Lagrangian for the particle in gravitational field $\varphi$: $\mathcal{L}%
$=$\dfrac{m\mathbf{v}^{2}}{2}-m\varphi.$ The Newton's equation for such a
Lagrangian reads:
\begin{equation}
\mathbf{\dot{v}=-\nabla}\varphi\mathbf{.} \tag{1}%
\end{equation}
Since the mass drops out from this equation, it is possible to think about
such an equation as an equation for a geodesic in (pseudo)Riemannian space.
\ This observation, indeed, had lead Einstein to the full development of
theory of general relativity. The above example is misleading though. Indeed,
let us consider the 2-body Kepler problem for particles with masses $m_{1}$
and $m_{2}$ interacting gravitationally. The Lagrangian for this problem is
given by%
\begin{equation}
\mathcal{L}=\frac{m_{1}}{2}\mathbf{\dot{r}}_{1}^{2}+\frac{m_{2}}%
{2}\mathbf{\dot{r}}_{2}^{2}+\gamma\frac{m_{1}m_{2}}{\left\vert \mathbf{r}%
_{1}-\mathbf{r}_{2}\right\vert }. \tag{2}%
\end{equation}
Introducing, as usual, the center of mass \ and relative coordinates via
$m_{1}\mathbf{r}_{1}+m_{2}\mathbf{r}_{2}=0$ and $\mathbf{r}=\mathbf{r}%
_{1}-\mathbf{r}_{2},$ the above Lagrangian \ acquires the form:%
\begin{equation}
\mathcal{L=}\frac{\mu}{2}\mathbf{\dot{r}}^{2}+\gamma\frac{m_{1}m_{2}%
}{\left\vert \mathbf{r}\right\vert }\equiv\frac{m_{1}m_{2}}{m_{1}+m_{2}}%
(\frac{\mathbf{\dot{r}}^{2}}{2}+\gamma\frac{(m_{1}+m_{2})}{\left\vert
\mathbf{r}\right\vert }), \tag{3}%
\end{equation}
where, as usual, we set $\mu=\frac{m_{1}m_{2}}{m_{1}+m_{2}}.$The constant
$\frac{m_{1}m_{2}}{m_{1}+m_{2}}$ can be dropped and, after that, instead of
the geodesic Eq. (1) we obtain the equation for a fictitious point-like object
of unit mass moving in the field of gravity produced by the point-like body of
mass $m_{1}+m_{2}$. Clearly, \ if $m_{1}$ is, say, the mass of the Sun, then
for different $m_{2}^{\prime}s$, one cannot talk about the geodesics.
Nevertheless, Infeld and Schild attempted to do just this in 1949 [$23$].\ The
case is far from being closed in 2015 [$24$]. These efforts look to us mainly
as academic (unless dynamics of binary stars is considered) for the following
reasons. If, say, $m_{1}\gg m_{2\text{ }}$ as for the electron in Hydrogen
atom or for the Mercury rotating around the Sun \ one can (to a very good
accuracy) discard mass $m_{2\text{ }}$ thus obtaining the equation for a
geodesic coinciding with Eq. (1). In the Introduction we defined the ratio
$r=\frac{m_{2}}{m_{1}+m_{2}}.$ If we do not consult reality for a guidance,
the ratio $r$ can have \textsl{any} nonnegative value. However, what is
observed in the sky (and in the atomic systems as well) leads us to the
conclusion that (excluding our Moon) all satellites of heavy planets as well
as all planets of our Solar System move along geodesics described by Eq. (1),
provided that we can ignore the interaction between the planets/satellites. We
shall call such an approximation the \textsl{Einsteinian limit, although it
was actually known already to Poincar}$e^{\prime}$\textsl{ }[$25$]\textsl{. It
is exactly equivalent to the mean field Hartree-type approximation in atomic
mechanics}. If we believe Einstein, then such Hartree-type approximation does
not require any corrections. This looks like ``too good to be true''. Indeed,
the first who actually exploited Einstein's limit (more then 100 years before
Einstein, and not realizing that this limit  is now known as the Einstein limit!) in
his calculations was Laplace [$26$]\textbf{,} Vol.4. \ In his book [25],
Vol.1, article 50, Poincare$^{^{\prime}}$ discusses Laplace's work on dynamics
of satellites of Jupiter. Quoting from Poincare$^{^{\prime}}$:

``(Following Laplace) consider the central body of large mass \ (Jupiter) and
three other small bodies (satellites Io, Europe and Ganymede), \textsl{whose
masses} \textsl{can be taken to be zero, }rotating around a large body in
accordance with Kepler's law. Assume further that the eccentricities and
inclinations of the orbits of these (zero mass) bodies are equal to zero, so
that the motion is going to be circular. Assume further that the frequencies
of their rotation $\omega_{1},\omega_{2}$ and $\omega_{3\text{ }}$are such
that there is a linear relationship%
\begin{equation}
\alpha\omega_{1}+\beta\omega_{2}+\gamma\omega_{3\text{ }}=0 \tag{4}%
\end{equation}
with $\alpha,\beta$ and $\gamma$ being three mutually simple integers such
that
\begin{equation}
\alpha+\beta+\gamma=0. \tag{5}%
\end{equation}
Given this, it is possible to find another three integers $\lambda
,\lambda^{\prime}$ and $\lambda^{\prime\prime}$ such that $\alpha\lambda
+\beta\lambda^{\prime}+\gamma\lambda^{\prime\prime}=0$ implying that
$\omega_{1}=\lambda A+B,\omega_{2}=\lambda^{\prime}A+B,\omega_{3}%
=\lambda^{\prime\prime}A+B$ with $A$ and $B$ being some constants. After some
time $T$ \ it is useful to construct the angles \ $T(\lambda A+B),T(\lambda
^{\prime}A+B)$ and $T(\lambda^{\prime\prime}A+B)$ describing current location
of respective satellites (along their circular orbits) and, their differences:
$(\lambda-\lambda^{\prime})AT$ and $(\lambda-\lambda^{\prime\prime})AT.$ If
now we choose $T$ in such a way that $AT$ is proportional to $2\pi,$ then the
angles made by the radius-vectors (from central body to the location of the
planet) will coincide with those for $T=0$. Naturally, such a motion (with
zero satellite masses) is periodic with period $T$.

\textsl{The question remains: Will the motion remain periodic in the case if
masses are small but not exactly zero? That is, if one allows the satellites
to interact with each other?.}...

\textsl{Laplace demonstrated that the orbits of these three satellites of
Jupiter will differ only slightly from truly periodic. In fact, the locations
of these satellites are oscillating around the zero mass trajectory}''

Clearly such an oscillation is taking place in such a way that the actual
(physical) orbits remain closed. In terminology of modern mathematical
literature [$27$] these orbits are known as the \textsl{Laplace-Lagrange}
(oscillating) orbits\footnote{E.g. make a Google search by typing in
``Laplace-Lagrange orbits''}. Evidently, the snapshot of such closed orbits is
looking as a standing wave, analogous to that for a particle in the box or for
an electron on the Bohr orbit. If this is so, can this fact be indicative of
the quantum nature of these orbits? The attentive reader of this excerpt from
Poincar$e^{\prime}$probably already noticed that the key ingredient in this
chain of reasonings is Eq.(4). Thus, it remains to prove that the condition,
Eq. (4), can be called the \textsl{quantization condition. If this is possible
to achieve, then it follows that only motions on Einsteinian trajectories is
compatible with Bohr-Sommerfel'd type \ quantization condition. That is at the
scales of Solar System}\textbf{ }\textsl{correctness of Einsteinian general
relativity is assured by correctness of quantum}\textbf{ }\textsl{mechanics
and vice versa.}

The quantization condition, Eq.(4) was chosen by Heisenberg [$28$] as
fundamental quantization condition from which all \ machinery of quantum
mechanics can be deduced! \textsl{Both in Solar system and in quantum
mechanics this type of conditions originate from the analysis of experimental
data: e.g. the astronomically observed rotation frequencies }$\omega
_{1},\omega_{2}$\textsl{ and }$\omega_{3\text{ }}$\textsl{ for satellites and}
\textsl{spectroscopically observed differences between frequencies} (e.g. see
(7a) below) of rotation of electrons on stationary Bohr orbits. This topic is
\ discussed further in the next subsection. Before doing so, we notice that
extension of work by Laplace to the full $n+1$ body planar problem was made
only in 20th century and can be found in the monograph by Charlier [$29$].
More rigorous mathematical proofs involving KAM theory have been obtained just
recently by Fejoz [$30$] and Biasco et al [$31$]. The difficulty, of course,
is caused by the proper accounting of the effects of finite but nonzero masses
of satellites and \ by showing that, when these masses are small, the
Einsteinian limit makes perfect sense and is stable. \ A sketch of these
calculations for planar four-body problem (incidentally studied by de Sitter
in 1909!) is presented in a nicely written lecture notes by Moser and Zehnder
[$32$].\medskip

\subsection{From Laplace to Heisenberg and Beyond}

We begin with the observation that the Schr\"{o}dinger equation cannot be
reduced to \ a simpler equation related to our macroscopic experience.
\textsl{It has to} \textsl{be postulated}.\footnote{Usually used appeal to the
DeBroigle wave-particle duality is of no help since the wave function in the
Schr\"{o}dinger's equation plays an auxiliary role.} On the contrary,
Heisenberg's basic equation \ from which quantum mechanics can be recovered is
inseparably linked with the experimental data and looks almost trivial.
Indeed, following Bohr, Heisenberg looked at the famous equations for energy
levels difference%
\begin{equation}
\omega(n,n-\alpha)=\frac{1}{\hbar}(E(n)-E(n-\alpha)), \tag{6}%
\end{equation}
where both $n$ and $n-\alpha$ are some integers. He noticed [$28$] that this
definition leads to the following fundamental composition law:%

\begin{equation}
\omega(n-\beta,n-\alpha-\beta)+\omega(n,n-\beta)=\omega(n,n-\alpha-\beta).
\tag{7a}%
\end{equation}
Since by design $\omega(k,n)=-\omega(n,k),$ the above equation can be
rewritten in a symmetric form as
\begin{equation}
\omega(n,m)+\omega(m,k)+\omega(k,n)=0. \tag{7b}%
\end{equation}
In such a form it is known as the honeycomb equation (condition) in current
mathematics literature [$34-36$] where it was rediscovered entirely
independently of \ Heisenberg's \ key quantum mechanical paper\ and,
apparently, with different purposes in mind. Connections between
\ mathematical results of Knutson and Tao [$34-36$] and those of Heisenberg
were noticed and developed in recent papers by Kholodenko [$37,38$]. We would
like to use some results from these works now. It should be noted though that,
unlike Schr\"{o}dinger's paper, whose results are discussed in any textbook on
quantum mechanics, the key Heisenberg's work [$28$] still had not found its
place in the textbooks. To fix this deficiency, and attempt was made recently
in [$39$]. \ In spite of this, to our knowledge, Heisenberg's paper is largely
unfamiliar among physics educated readers, perhaps, because of the fact that
technically Schr\"{o}dinger's approach to quantum mechanics superficially is
looking much simpler and because it is widely believed that both approaches
are equivalent.

We begin our discussion of Heisenberg's ideas by noticing that Eq.(7b) due to its
purely combinatorial origin does not contain the Plank's constant $\hbar$.
Such fact is of major importance for this work since the condition Eq. (4) can
be equivalently rewritten in the form of Eq.(7b), where $\omega(n,m)=\omega
_{n}-\omega_{m}$. It would be quite unnatural to think of the Planck's
constant in this case.

Eq. (7b) looks almost trivial and yet, it is sufficient for restoration of all
quantum mechanics. Indeed, in his paper of October 7th of 1925, Dirac [$40$],
\ being aware of Heisenberg's key paper\footnote{This paper was sent to Dirac
by Heisenberg himself prior to its publication.}, streamlined Heisenberg's
results and \ introduced \ notations which are in use up to this day. He
noticed that the combinatorial law given by Eq.(7a) for frequencies, when used
in the Fourier expansions for composition of observables, leads to the
multiplication rule $a(nm)b(mk)=ab(nk)$ for the Fourier amplitudes for these
observables. In general, in accord with Heisenberg's assumptions, one expects
that $ab(nk)\neq ba(nk).$ Such a multiplication rule is typical for matrices.
In traditional quantum mechanical language such matrix elements are written as
$<n\mid\hat{O}\mid m>\exp(i\omega(n,m)t)$ so that Eq.(7.b) is equivalent to
the matrix statement%
\begin{align}%
{\textstyle\sum\nolimits_{m}}
&  <n\mid\hat{O}_{1}\mid m><m\mid\hat{O}_{2}\mid k>\exp(i\omega(n,m)t)\exp
(i\omega(m,k)t)\nonumber\\
&  =<n\mid\hat{O}_{1}\hat{O}_{2}\mid k>\exp(i\omega(n,k)t). \tag{8}%
\end{align}
for some operator (observables) $\hat{O}_{1}$ and $\hat{O}_{2}$ evolving
according to the rule: $\hat{O}_{k}(t)=U\hat{O}_{k}U^{-1},k=1,2,$ provided
that $U^{-1}=\exp(-i\frac{\hat{H}}{\hbar}t).$ \ From here it follows that
$U^{-1}\mid m>=\exp(-\frac{E_{m}}{\hbar}t)\mid m>$ \ if one identifies
$\hat{H}$ with the Hamiltonian operator. Clearly, upon such an identification
the Schr\"{o}dinger equation can be obtained at once as is well known [$41$]
and, with it, the rest of quantum mechanics. In view of [$34-38$] it is
possible to extend the traditional pathway: from classical to quantum
mechanics and back. \ We begin discussion of this topic in the next section
and shall continue doing so from other perspectives in the rest of this
paper.\bigskip

\section{Newtonian Limit of Einsteinian   Gravity   and Quantum\\ Dynamics of  
Solar System (Specifics)}

\subsection{A Sketch of LeVerrier's Calculations}

After suggestive discussion of previous section it is instructive to arrive at
the same conclusions via entirely different \ route.\ \ This task is
accomplished in this section.\ \ In developing his theory of general
relativity Einstein was aware of previous efforts at calculation of bending of
light and perihelion shift for Mercury based on Newton's mechanics. In
particular, bending of light based on Newtonian theory was calculated by
Soldner in 1804 [$7,8$]. Calculation of the perihelion of Mercury, as well as
other planets was done by Le Verier [$8$] in 1859\footnote{E.g. see Wikipedia
\par
http://en.wikipedia.org/wiki/Urbain\_Le\_Verrier}. His computations yielded
526.7 seconds of arc per century as compared with \ the best modern
theoretical value of 532. Our readers are advised to read [$42$] before they
continue with reading. The observed precession is known to be 575. The
difference between these two is 43 seconds of arc per century. It is famous
Einstein's result for the perihelion shift for Mercury. In Einstein's
calculations no mention of Le Verrier's results were made. This leaves us with
the following problem. If we believe Einstein, then Mercury should move on a
geodesic. But if this is so, then how to look at Le Verrier's result obtained
by explicit account of \ gravitational interactions between the Mercury and
the rest of planets? This brings us back to results of previous section. This
time, however, we would like to look at the same problem differently. For this
purpose, we are going to discuss briefly the results by Le Verrier in a
simplified form taken from [$8$].

\medskip Taking into account the planarity of motion in the gravitational
field \ we introduce the polar coordinates $r(t)$ and $\varphi(t)$ of the
particle of unit mass (e.g. see Eq. (3)) moving in the gravity field of
massive body of mass $M.$ Newton's equations are given by\footnote{In the
system of units in which the gravitational constant $\gamma$ was put equal to
one.}
\begin{equation}
\ddot{r}-r\dot{\varphi}^{2}=-\frac{M}{r^{2}}\text{ and }r\ddot{\varphi}%
+2\dot{r}\dot{\varphi}=0 \tag{9}%
\end{equation}
implying conservation of \ the angular momentum $L=r^{2}\dot{\varphi}$ and
allowing the first of Eq.s (9) to be rewritten in the form%
\begin{equation}
\ddot{r}-\frac{L^{2}}{r^{3}}=-\frac{M}{r^{2}}. \tag{10}%
\end{equation}
Taking into account that $\dot{\varphi}=\frac{L}{r^{2}}$ the following chain
of transformations%
\begin{equation}
\dot{r}=\left(  \frac{d\varphi}{dt}\frac{dt}{d\varphi}\right)  \frac{dr}%
{dt}=\frac{L^{2}}{r^{2}}\frac{dr}{d\varphi} \tag{11}%
\end{equation}
and%
\begin{equation}
\ddot{r}=L\left(  \frac{d\varphi}{dt}\frac{dt}{d\varphi}\right)  \frac{d}%
{dt}(r^{-2}\frac{dr}{d\varphi})=\frac{L^{2}}{r^{2}}\frac{d}{d\varphi}%
(r^{-2}\frac{dr}{d\varphi}) \tag{12}%
\end{equation}
is useful. Substituting this result in Eq.(10) produces%
\begin{equation}
\frac{d}{d\varphi}(r^{-2}\frac{dr}{d\varphi})-\frac{1}{r}=-\frac{M}{L^{2}}.
\tag{13}%
\end{equation}
Let now $u=\frac{1}{r},$ then the above equation acquires especially simple
form
\begin{equation}
\frac{d^{2}u}{d\theta^{2}}+u=\frac{M}{L^{2}}. \tag{14}%
\end{equation}
The constant term on the r.h.s can be easily eliminated so that we are left
with the equation of motion for the harmonic oscillator. Consider now the
related, more general, equation%
\begin{equation}
\frac{1}{\Omega^{2}}\frac{d^{2}u}{d\varphi^{2}}+u=\frac{1}{P} \tag{15}%
\end{equation}
where both $\Omega^{2}$ and $P$ are some constants. \ This equation admits\ a
solution%
\begin{equation}
u(\varphi)=\frac{(1+k\cos(\Omega\varphi))}{P} \tag{16}%
\end{equation}
in which $k$ is a constant of integration. This result can be converted into
polar equation for an ellipse. In the case of Eq.(14) it is given by
\begin{equation}
r(\varphi)=\frac{L^{2}}{M}\frac{1}{1+k\cos\varphi}. \tag{17}%
\end{equation}
We would like now to complicate this situation as follows. Suppose that
obtained results, say, \ are for the Mercury. But then, in its current form
they are unrealistic since Mercury is interacting only with the Sun. To
account for interactions of Mercury with other planets requires much more
work. This was done by Le Verrier and takes about 150 pages\footnote{E.g.see
footnote 11 and Note added in proof.} . Fortunately, there is a much easier
method known as the mass ring model described in [$8$] which produces results
very close to those by Le Verrier. The idea of the method lies in replacement
of the planets other than Mercury by gravitating rings \ of masses $m_{i}$
$(i=1-8)$ centered at the Sun \ and located further away from the Sun\ at the
radial distance $R_{i}$.\ \ In such a case Mercury will be experiencing the
inward force coming from the Sun and the outward force coming from rings. The
outward potential $\psi(r)$ of a particular ring at the distance $r$ \ from
the Sun \ and in the plane of a ring (which is the same for all planets) is
given by\footnote{We drop the subscript i for brevity.}
\begin{equation}
\psi(r)=-\frac{m}{R}[1+\frac{1}{4}\left(  \frac{r}{R}\right)  ^{2}+\frac
{9}{64}\left(  \frac{r}{R}\right)  ^{4}+\frac{25}{256}\left(  \frac{r}%
{R}\right)  ^{6}+...]. \tag{18}%
\end{equation}
In view of this result, the Newton's equation of motion, Eq.(10), is modified
in the presence of a ring as follows%
\begin{equation}
\ddot{r}-\frac{L^{2}}{r^{3}}=-\frac{M}{r^{2}}+\alpha_{1}r+\alpha_{2}%
r^{3}+\cdot\cdot\cdot\tag{19}%
\end{equation}
with $\alpha_{1}=\frac{m}{2R^{3}},$ $\alpha_{2}=\frac{9m}{16R^{5}},$ and so
on. Let $r_{1}$ and $r_{2}$ be the minimum and maximum radial distances from
the Sun for Mercury. These numbers can be used for simplification of Eq.(19).
Indeed, it can be brought into the form%
\begin{equation}
\ddot{r}-\frac{L^{2}}{r^{3}}=-\frac{A}{r^{2}}-\frac{B}{r^{3}}, \tag{20}%
\end{equation}
provided that the constants $A$ and $B$ are determined from the equations%
\begin{equation}
\frac{A}{r_{1}^{2}}+\frac{B}{r_{1}^{3}}=\frac{M}{r_{1}^{2}}-\alpha_{1}%
r_{1}-\alpha_{2}r_{1}^{3}-\cdot\cdot\cdot\tag{21a}%
\end{equation}
and%
\begin{equation}
\frac{A}{r_{2}^{2}}+\frac{B}{r_{2}^{3}}=\frac{M}{r_{2}^{2}}-\alpha_{1}%
r_{2}-\alpha_{2}r_{2}^{3}-\cdot\cdot\cdot. \tag{21b}%
\end{equation}
If \ $r_{1}$ differs not much from $r_{2}$ (small eccentricity) it is possible
to introduce the mean orbital radius \ $r_{0}=\frac{1}{2}(r_{1}+r_{2})$ so
that the constants $A$ and $B$ can be represented by the following power
series expansions%
\begin{equation}
A=M-4\alpha_{1}r_{0}^{3}-6\alpha_{2}r_{0}^{5}-... \tag{22a}%
\end{equation}
and%
\begin{equation}
B=\text{ \ \ \ }3\alpha_{1}r_{0}^{4}+5\alpha_{2}r_{0}^{6}+\cdot\cdot\cdot.
\tag{22b}%
\end{equation}
Using these results, Eq.(20) can now be rewritten in the form of Eq.(14) (or
Eq.(15)). This can be done as follows. By writing Eq.(20) as%
\[
\ddot{r}-\left(  \frac{L^{2}-B}{r^{3}}\right)  =-\frac{A}{r^{2}}%
\]
and by assuming that the angular momentum $L$ is conserved it is convenient to
rewrite the above equation as follows%
\[
\frac{1}{1-\frac{B}{L^{2}}}\ddot{r}-\frac{L^{2}}{r^{3}}=-\frac{A}{r^{2}}%
\frac{1}{1-\frac{B}{L^{2}}}.
\]
For this equation we can repeat the same steps as lead from Eq.(10) to Eq.(14)
in order to obtain the equation analogous to Eq.(15) in which $\Omega
=\sqrt{1-\frac{B}{L^{2}}}$ and $P=\left(  1-\frac{B}{L^{2}}\right)  /A.$
Elementary calculation,\ e.g. see [$21$], produces: $L^{2}=Mr_{0}$. Using this
result we obtain,%
\begin{equation}
\Omega=\sqrt{1-\frac{B}{L^{2}}}\simeq1-\frac{1}{2}\frac{B}{Mr_{0}}. \tag{23}%
\end{equation}
It can be shown using this result for $\Omega$ [$8$] that contribution of the
particular ring to the Newtonian shift \ is given by%
\begin{align}
\Delta\varphi &  =\frac{\pi B}{Mr_{0}}\nonumber\\
&  =\pi\left(  \frac{m}{M}\right)  [\frac{3}{2}\left(  \frac{r_{0}}{R}\right)
^{3}+\frac{45}{16}\left(  \frac{r_{0}}{R}\right)  ^{5}+\cdot\cdot\cdot].
\tag{24}%
\end{align}
To use this result for calculation of the perihelion shift for Mercury, we
have to add up contributions, e.g. numerical values for masses \ and mean
radiuses, for all rings/planets (excluding Pluto and including the asteroid
belt). The obtained result is 549.7 \ seconds of arc per century compares well
with Le Verrier's 526.7. \ We shall analyse these results further in the next subsection

\subsection{Sketch of Einstein-type Calculations} 

In [$43$] Le Verrier's result for the perihelion shift for Mercury was
characterized as ``the first relativistic gravity effect observed''. Based on
results of the preceding subsection this statement is incorrect. Such a
conclusion is in accord with commonly accepted \ point of view, e.g. read
Chr.9, paragraph 5 of [$44$]. The situation however is more delicate as it
appears. In this and the following subsection we would like to discuss why
this is so. \ 

Following Ref.[$8$], in Einstein's case the equation of motion replacing
Eq.(10) is given by $(c=1)$%
\begin{equation}
\ddot{r}-\frac{L^{2}}{r^{3}}=-\frac{M}{r^{2}}-\frac{3ML^{2}}{r^{4}} \tag{25}%
\end{equation}
Superficially, it differs from Eq.(10) only by the presence of the last term.
However, in Einstein's case the evolution is taking place not in Newton's
absolute time but in \textsl{proper} time [$8,44$]. \ Since this fact does not
change the mathematical treatment of such an equation, we can go trough the
same steps as in previous subsection. That is, instead of Eq.(14), this time
we obtain%
\begin{equation}
\frac{d^{2}u}{d\varphi^{2}}+u=\frac{M}{L^{2}}+3Mu^{2}. \tag{26}%
\end{equation}
Presence of the last term causes this equation to be nonlinear so that its
exact solution, in principle, is much harder to find. To obtain Einstein's
result for the perihelion shift of the Mercury, it is sufficient to use just a
perturbation theory. For this purpose, again, following [$8$], we rewrite
Eq.(26) in the equivalent form%
\begin{align}
u  &  =\frac{1}{6m}(1-\sqrt{1-12(\frac{M^{2}}{L^{2}}-M\ddot{u})})\nonumber\\
&  \simeq\frac{1}{6m}[\frac{1}{2}(12M(\frac{M}{L^{2}}-\ddot{u}))+\frac{1}%
{8}(12M(\frac{M}{L^{2}}-\ddot{u}))^{2}+\cdot\cdot\cdot]. \tag{27}%
\end{align}
Rearranging terms in this result brings us back to Eq.(15), in which
$\Omega^{-1}=\sqrt{1+6\left(  \frac{M}{L}\right)  ^{2}}$, and $P^{-1}=$
$\frac{M}{L^{2}}+3\frac{M^{3}}{L^{4}}.$ In arriving at this result, in accord
with [$8$], extra terms containing $3M\ddot{u}^{2}$ were dropped. This is
legitimate because of the following. \ From observational astronomy it is
known that all three Kepler's laws hold to a large degree of accuracy for
planets of Solar system. This means that Eq.(15) is sufficiently adequate for
description of planetary motion. But if this is correct, then the perturbation
coming from the term $3M\ddot{u}^{2}$ should be negligible. Thus, we arrive at
the following \ relativistic analog of Eq.(16)
\begin{equation}
r(\varphi)=[\frac{L^{2}/M}{1+3\left(  \frac{M}{L}\right)  ^{2}}]\frac
{1}{1+k\cos(\Omega\varphi)}. \tag{28}%
\end{equation}
If $\Omega$ would be equal to one, the above equation is converted into
standard equation for an ellipse with the origin at one focus and the
eccentricity $k$. But since $\Omega<1$ the angle $\varphi$ must go beyond
$2\pi$ in order \ for the radial distance to complete one cycle around the
ellipse. This is being interpreted as precession. In the present case, the
precession angle per revolution is $\Delta\varphi=$ $2\pi(\frac{1}{\Omega
}-1).$ Since
\[
\Omega\approx1-3\left(  \frac{M}{L}\right)  ^{2}+\frac{27}{2}\left(  \frac
{M}{L}\right)  ^{4}+\cdot\cdot\cdot
\]
and expecting that $\left(  \frac{M}{L}\right)  \ll1,$ we obtain celebrated
Einstein's \ result for perihelion shift of Mercury [$45$]%
\begin{equation}
\Delta\varphi\simeq6\pi\left(  \frac{M}{L}\right)  ^{2}. \tag{29}%
\end{equation}
\bigskip

\subsection{From Analysis to Synthesis}

In this subsection we connect results obtained above, in this section, with
those in Section 2. Following Weinberg [$44$] (Ch.r 8, paragraph 6 ), we
notice that among all tests of general relativity the measurement of the
perihelion shift (say, for the Mercury) is the most important one. Such a
measurement cannot be done directly though. In practice what is being measured
is the total shift $\Delta\varphi_{obs}.$ Then, the Newtonian portion (e.g.
calculated in subsection 3.1) \textsl{is subtracted from} $\Delta\varphi_{obs}$
(e.g. see Eq.(8.6.13) of Ref. [$44$]). The remainder is the famous Einsteinian
part of the shift theoretically calculated in subsection 3.2. Although
formally such a procedure makes sense, it is not in accord with results of
Section 2. In Section 2 the Einsteinian limit was defined in which
planets/satellites are not interacting with each other (i.e. this is the
massless limit). Interaction effects are taken into account by considering
stability of the Einsteinian orbits (geodesics) against small gravitational
perturbations caused by interplanet/intersatellite interactions. These
interactions caused Einsteinian orbits to become the Laplace-Lagrange
oscillating orbits. These are analogs of\ standing waves for electrons in
Bohr's model of atom as we noticed already. The above computational scheme
uses the Newtonian mechanics in which the Newtonian Eq.(1) is identified with
the equation for Einsteinian geodesic. Such an identification requires some
care as explained in detail in [$44$] (Ch.r 8, paragraph 4). The difficulty
stems from the fact that in Einsteinian theory gravitational interactions are
propagating with finite speed while in Newtonian gravity the speed is
infinite. This causes in Einsteinian theory to use the \textsl{proper time.}
It is surely not the same as Newtonian time. To match these two theories in
the limit of small velocities it is instructive to discuss the compatibility
of Einsteinian dynamics with Keplerian laws. \ Fortunately, this issue is
considered in detail in [$8$], (Chr.5.5). More on this will be presented
below, in sections 4- 6. This allows us to reduce our discussion to the
absolute minimum.

The fact that Newton's law of gravity was deduced from Kepler's laws is well
known. Much less is known that the Schwarzschield metric can \ also be
restored using the 3rd Kepler's law. To write this law explicitly requires few
steps which would be unnecessary should the recommended textbooks contain
needed information. In these textbooks the Lagrangian $\mathcal{L}$ for a
particle of mass m$_{2}$ interacting with another particle of mass m$_{1}$ is
given by Eq.(2) or, after usual reduction to center of mass coordinates, by
\begin{equation}
\mathcal{L=}\frac{\mu}{2}\mathbf{\dot{r}}^{2}+\gamma\frac{m_{1}m_{2}%
}{\left\vert \mathbf{r}\right\vert }. \tag{30}%
\end{equation}
In such a form it is given, for example, in [$21$]. For our purposes it is
more advantageous to use another form of the Lagrangian
\begin{equation}
\mathcal{L=}\frac{\mathbf{\dot{r}}^{2}}{2}+\gamma\frac{M}{\left\vert
\mathbf{r}\right\vert } \tag{31}%
\end{equation}
also given in Eq.(3). Evidently, in Eq.(31) $M\simeq m_{1}.$ If $T$ is period
of revolution of the fictitious particle of unit mass around the origin of
coordinate system centered at $M$, then the 3rd Kepler's law \ reads
($\gamma=1$ as before)
\begin{equation}
T=2\pi r_{0}^{\frac{3}{2}}\sqrt{\frac{1}{M}}\text{ or }\omega^{2}r_{0}^{3}=M.
\tag{32}%
\end{equation}
In this formula we used $r_{0}=\frac{1}{2}(r_{1}+r_{2}),$ as before, and
$\omega=2\pi/T$. \ But, because $\omega=\dot{\varphi}=\frac{d}{dt}\varphi,$ it
is permissible to write as well $\omega=\frac{d}{dt}\varphi=\frac{d}{d\tau
}\varphi\left(  \frac{d\tau}{dt}\right),$ where $\tau$ is some function of
\ Newton's time $t$. Now we would like to identify $\tau$ with the
\textsl{proper time} of Einsteinian gravity. This is needed \ in view of the following.

Recall that in polar coordinates Einstein's equation of motion for fictitious
particle of unit mass is given by Eq.(25). It differs from Newtonian's Eq.(10)
in two ways. First, instead of the Newtonian absolute time, in the present
case the differentiation is taking place over the \textsl{proper time}.
Second, there is an extra potential term absent in Newtonian version. This
difference between the Newtonian and Einsteinian \ formulations of this
mechanical problem is only apparent. It can be eliminated with the help of the
3rd Kepler's law, Eq.(32), as explained in [$8$]. \ Since this explanation is
scattered allover this reference, we would like to describe it now in a
concise form. Even without going into fine details, the results of previous
subsections indicate that \ the Einsteinian result for the perihelion shift of
Mercury can be \ formally obtained simply by readjusting the Newtonian
potential(s). Incidentally, this was the original Le Verier's idea \ since he
believed that between Mercury and Sun there should be yet another planet
(Vulcan) which only remains to be discovered. Hence, such an homomorphism
between the Newtonian and Einsteinian mechanics of point particles is assured
from this point of view. Nevertheless, as is well known, absence of Vulcan had
brought to life general relativity. It is very instructive to supply the
derivation \ of this homomorphism explicitly. For this purpose we take into
account that for the particle of unit mass $L=\omega r^{2}$. Using this
result, Eq.(25) can be rewritten as
\begin{equation}
\ddot{r}=-\frac{M}{r^{2}}+\omega^{2}r(1-\frac{3M}{r}). \tag{33}%
\end{equation}
To analyze this equation further we need to recall that for a non-rotating
body the Scwarzschield metric has the form%
\begin{equation}
\left(  d\tau\right)  ^{2}=[1-\frac{2M}{r}]\left(  dt\right)  ^{2}-[\frac
{1}{1-\frac{2M}{r}}]\left(  dr\right)  ^{2}-r^{2}\left(  d\theta\right)
^{2}-r^{2}\sin^{2}\theta\left(  d\phi\right)  ^{2}. \tag{34}%
\end{equation}
Consider a special case of this metric for a motion in the equatorial plane
$\theta=\frac{\pi}{2}$ taking place on a circle of radius $R$. In such a case
the metric can be rewritten in polar coordinates as
\begin{equation}
\left(  d\tau\right)  ^{2}=[1-\frac{2M}{R}]\left(  dt\right)  ^{2}%
-R^{2}\left(  d\varphi\right)  ^{2}, \tag{35}%
\end{equation}
where we have relabeled $\phi\rightleftarrows\varphi$ for consistency with
previous notations. If we treat $R$ as $r_{0}$ (defined in Eq.(32)), then we
can use the 3rd Kepler's law in order to write%
\begin{equation}
R^{2}=M/(\omega^{2}R)=\left(  \frac{dt}{d\varphi}\right)  ^{2}\frac{M}{R}.
\tag{36}%
\end{equation}
Upon substitution of this result into Eq.(35) we obtain,%
\begin{equation}
\left(  \frac{d\tau}{dt}\right)  ^{2}=1-\frac{3M}{R}. \tag{37}%
\end{equation}
This result can be used in Eq.(33) (\textsl{evidently for closed orbits only
with not too large} \textsl{eccentricity}). Then, using Eq.(37) we obtain,%
\begin{equation}
\omega^{2}(1-\frac{3M}{r})=\left[  \frac{d}{d\tau}\varphi\left(  \frac{d\tau
}{dt}\right)  \right]  ^{2}=\tilde{\omega}^{2}, \tag{38}%
\end{equation}
where we relabeled $R\rightarrow r_{0}\rightarrow r$ and used the fact that
$\ \omega=\dot{\varphi}=\frac{d}{dt}\varphi$ is the time derivative of
$\varphi$ with respect to \textsl{the Schwarzschield coordinate time}. This
result allows us\ to rewrite the Einsteinian Eq.(33) in the Newtonian form%
\begin{equation}
\ddot{r}-\tilde{\omega}^{2}r=-\frac{M}{r^{2}} \tag{39}%
\end{equation}
\ formally coinciding with Eq.(9) since\ we already have established that
$\omega=\frac{d}{dt}\varphi=\frac{d}{d\tau}\varphi\left(  \frac{d\tau}%
{dt}\right)  $.

The obtained result is only a homomorphism (not an isomorphism) in view of the
following. Eq.(39) underscores the non triviality of time transformations.
These are normally not present in Newtonian mechanics. In particular, we
notice that for $\hat{R}=3M$ in Eq.(37) $\frac{d\tau}{dt}=0.$ This result is
of \ physical significance. It describes the radius of the photon sphere.
\ This result can be \ easily derived by noticing that in the case of light
Eq.(26) acquires the form [$8$]%
\begin{equation}
\frac{d^{2}u}{d\varphi^{2}}+u=3Mu^{2} \tag{40}%
\end{equation}
formally implying that for photons $L\rightarrow\infty.$ Both Eq.(26) and (40)
possess a fixed point solution $u^{\ast}=const$ \ implying that either
\begin{equation}
u^{\ast}=\frac{M}{L^{2}}+3Mu^{\ast2}\text{ or }u^{\ast}=3Mu^{\ast2}, \tag{41}%
\end{equation}
provided that%
\begin{equation}
\frac{d}{d\tau}\varphi=Lu^{\ast2}\text{ since }L=\omega r^{2}. \tag{42}%
\end{equation}
In the ``massive'' case \ we have
\[
L^{2}(u^{\ast}-3Mu^{\ast2})=M\text{ \ }%
\]
so that
\begin{equation}
\left(  \frac{d}{d\tau}\varphi\right)  ^{2}=\frac{M}{r^{\ast2}(r^{\ast}-3M)}.
\tag{43}%
\end{equation}
The positivity of the l.h.s is implying that $r^{\ast}-3M>0.$ That is for
massive particles all circular orbits should have radius $r>3M.$ The radius
$r^{\ast}=3M$, e.g. see Eq.(41), is possible only for photons and describes
the photon sphere. Consider now the stability of \ the photon sphere. It can
be easily calculated using the following representation of Eq.(40)%
\begin{equation}
\frac{du}{d\varphi}=p \tag{44a}%
\end{equation}
and%
\begin{equation}
\frac{dp}{d\varphi}=u(3Mu-1). \tag{44b}%
\end{equation}

For $u^{\ast}$ such that $3Mu^{\ast}=1$ standard stability analysis produces
eigenvalues $\varepsilon=\pm1$ for the stability matrix indicating that
\textsl{classically} the photon sphere orbits are unstable. Since photons are
quantum objects such classical analysis may or may not be valid. Thus, in the
case of photon sphere we are confronted with an unusual situation. While we
had began our analysis using classical mechanics (classical general
relativity), we ended up with the necessity to discuss quantum mechanical
effects in general relativity. Such situation is not totally unexpected in
view of recent attempts to use the formalism of general relativity for
designing of variety of photo devices in laboratory conditions [$11-13$].
Whether in \ real world or in laboratory, the progress in understanding of
general relativity (and, subsequently, of quantum effects in this theory)
begins with understanding of the difference between Eq.s(1) and (3) (or
between Lagrangians given by Eq.s (30) and (31)). This has been recognized by
Infel'd and Schild who attempted to analyze \ this difference already in 1949
[$23$].\ The case is far from being closed \ even today [$24$].

\ The decisive attempt to describe the motion of extended objects in general
relativity was made in seminal paper by Papapetrou [$46$] and continues up to
the present day.\ From his papers it is known that, strictly speaking, the
motion of extended bodies is not taking place on geodesics. And yet, for the
Mercury such an approximation made by Einstein works extremely well as we just
\ have demonstrated. Hence, the main issue is to understand the difference
between the massless (photon) and the massive cases. Evidently, the
Lagrangian, Eq.(31), remains the same as long as the mass $m_{2}$ in Eq.(3) or
(31) is small but nonzero. In particular, \ this means that such a mass can be
made arbitrarily small since relativity in its original form does not impose
lover limit restrictions on the mass. Only the difference between the massive
and the massless cases matters. Indeed, to obtain the results for the
perihelion shift or bending of light nowhere in our calculations we had \ used
the mass $m_{2}$. \ There are treatments [$22$] involving such a mass (in the
post Newtonian approximation). But, \ since it drops out at the end of
calculations anyway, it is possible to perform all calculations \ without
using this mass from the beginning [$47$]. Then the difference between
different massive bodies disappears. \ Because of \ this, for the sake of
argument, we can replace Mercury by, say, the atomic nucleus, or, just a
single proton (or neutron), etc. In such a case not only photon sphere but the
rest of massive orbits become \ \textsl{formally} quantum. \ Evidently, if we
are able to apply this reasoning to Mercury, we\ can do the same for the rest
of planets in view of smallness of their masses as compared to that of the
Sun. This indeed was accomplished [$42,43,48,49$]. In such a case, we can
totally neglect their mutual attraction thus arriving at the situation
considered already by Laplace, improved by Lagrange, Poincare$^{\prime}$ and
others as discussed it in section 2 where we gave it \ the name
``\textsl{Einstein limit}''. But we \ mentioned \ in the same section that to
study the stability of Einsteinian orbits it is necessary to consider
perturbations of the massless orbits caused by interaction between masses
considered to be small. \ In the case of Mercury, the influence of other
planets was discussed in subsection 3.1. resulting in Eq.s(22) and (23). From
these equations it follows that influence of other planets causes us to
consider the Newtonian-type equations of motion in which the mass $M$ and
\ the square of angular momentum $L^{2}$ should be redefined \textsl{without
changing} \textsl{the form} of Newton's equations. This\ observation is in
accord with our earlier claim in subsection 2.1. that Einsteinian general
relativity could be looked upon as Hartree-type approximation in atomic physics.

Next, in this subsection we deonstrated that Einsteinian equations can be
formally brought to the Newtonian form. As we discussed at the beginning of
this subsection, the perihelion shift obtained by Einstein for Mercury is the
difference between $\Delta\varphi_{obs}$ and $\Delta\varphi_{Newton}$. \ In
view of previous remarks, this means that we can turn the above arguments
around and to say that both contributions to perihelion shift can be obtained
simultaneously from the Newton-looking (but actually the Einstein-type)
equation for geodesics in which the parameters $M$ and $L^{2}$ are carefully
chosen. Such a statement is formally compatible with the results of Laplace
\ discussed in section 2 and those stated in [$43$] but is still not complete.
It is not complete because we have not accounted yet for the relationship of
the type given by Eq.(4). To account for the relationship between angular
frequencies is nontrivial. Indeed, now we know that if we begin with Newton's
Eq.(39), perhaps with properly redefined parameters $M$ and $\omega,$ we can
always use it in order to re obtain Einstein's \ Eq.(33) for geodesics. But in
Einsteinian universe geodesics should be parametrized only once while in the
present case (\textsl{without invoking quantization!)} such parametrization is
apparently different for each planet other than Mercury. \ This happens
because \ LeVerier's contributions are redefining geodesic parameters for
different planets in apparently different and uncorrelated way. This apparent
contradiction can be eliminated by assuming that parameters of the lowest
possible stable geodesic orbit, say, for the Mercury determine parameters for
geodesics of other planets. This suggestion is compatible with the fact that
we can place a proton instead of the Mercury on the same geodesic as we
already mentioned. In such a case the problem about proton trajectory is no
longer lying only in the classical domain and, indeed, we shall demonstrate
below that \textsl{the parameters for higher orbits are determined by the
parameters of the first allowed orbit which in the massive case should play
the same role as photon sphere in the massless}. Such a conclusion is
compatible with results of Section 2 but requires explicit calculations for
demonstration of its correctness to be accomplished in the rest of this paper.
We hope that the results obtained below could be tested not only against the
empirical data in the sky (as it is done in this paper) but, subsequently, in
the laboratory [$11-13$].

\section{Space, Time and Space-Time in Classical and Quantum\\ Mechanics  and General Relativity. Bertrand Spacetimes}

\subsection{Space and Time in Classical Mechanics}

If one contemplates quantization of dynamics of celestial objects in a simple
minded fashion using standard textbook prescriptions, one immediately runs
into myriad of small and large problems. Unlike atomic systems in which all
electrons repel each other, have the same masses and \ are indistinguishable,
in the case of Solar system all planets (and satellites) attract each other,
have different masses and visibly distinguishable. Besides, in the case of
atomic systems the Planck constant $\hbar$ plays prominent role while no such
a role can (or should) be given to the Planck constant in the sky. This will
be explained below. Above we argued that in the Einsteinian limit it is
possible to remove almost all of these objections so that, apparently, the
only difference between the atomic and celestial mechanics (should we be ready
to quantize it) lies in replacement of the Planck constant by another (yet
unknown) constant and in attraction between masses (planets/satellites),
instead of repulsion, between electrons.

Although \ the celestial mechanics based on Newton's law of gravity \ is
apparently classical (i.e. non quantum), \ with such an assumption one easily
runs into serious problems. Indeed, this leaves completely unexplained such an
empirical fact \ that the typical ratio between masses of all planets of Solar
system and that of the Sun are about 10$^{-6},$ the same goes about masses of
satellites of heavy planets vs. masses of these planets. \textsl{Classical
mechanics does not impose any restrictions on masses}! Next, in Newton's
mechanics the speed of propagation of gravity is assumed to be infinite and
time is \ absolute as is well known. These properties were \textsl{
effectively postulated} by Newton. Whether or not this is true or false could
be decided only experimentally. As result, general relativity had emerged.
Since at the \ spatial scales of Solar system one has to use radio signals to
check correctness of Newton's mechanics, all kinds of wave mechanical effects
such as retardation, the Doppler effect, etc. become of use. Because of this,
measurements are necessarily having some error margins. The error margins
naturally will be larger for more distant objects. \textsl{Accordingly, even
at the level of classical Newtonian\ celestial mechanics we have to deal with
inaccuracies of measurements exactly analogous to those} \textsl{in atomic
mechanics. These probabilistic effects are unavoidable but are not taken into
account \ whatsoever in classical mechanics. Should they be taken into
account, the distiction between classical and quantum mechanics would
disappear. Furthermore, the common belief that quantum mechanics is
exclusively the domain of macrocosm happen to be false since it can be
demonstrated, e.g. read }[$33$], \textsl{Appendix A, that the Heisenberg group
could be obtainable based on purely macroscopic considerations and has the
same right to exist in macroscopic reality as, say, the Lorentz group, the
rotation group, etc}.

To make the formalisms of both the atomic and celestial mechanics
\ compatible, we have to think carefully about the space, time and space-time
transformations at the level of classical mechanics. We begin with the
observation that in Hamiltonian mechanics the Hamiltonian equations \textsl{by
design} remain invariant with respect to the canonical transformations. Now we
would like to complicate this familiar picture by investigating the
``canonical'' time changes in classical mechanics even though in reality these
transformations are not necessarily canonical. Fortunately, in the case \ when
these are the canonical time transformations the task is accomplished to a
large extent in monograph by Pars [$50$]. \ For the sake of space, we refer
our readers to pages 535-540 of this monograph for details.

\ In accord with Dirac [$51$]$,$ we believe, that good quantization procedure
should always begin with the Lagrangian formulation of mechanics. Hence, we
also begin with the Lagrangian $\mathcal{L=L(}\{q_{i}\},\{\dot{q}_{i}\})$. The
equations of motion can be written in the form of Newton's equations $\dot
{p}_{i}=F_{i},$ where the generalized momenta $p_{i}$ are given by
$p_{i}=\delta\mathcal{L}/\delta\dot{q}_{i}$ and the generalized forces $F_{i}$
by $F_{i}=-\delta\mathcal{L}/\delta q_{i}$ as usual. In the case if the total
energy $E$ is conserved, it is possible instead of ``real'' time $t$ to
introduce the fictitious time $\theta$ via relation $dt=u(\{q_{i}\})d\theta$
where the function $u(\{q_{i}\})$ is assumed to be nonnegative and is
sufficiently differentiable with respect to its arguments. At this point we
can inquire if Newton's equations can be written in terms of new time variable
so that they remain form- invariant. \ We shall treat these equations as
Lagrangian equations. In such a case we do not need to worry about their
validity in relativistic domain \ since the Lagrangian mechanics in being used
in this domain as well. \textsl{From results \ discussed in previous section
it follows that use of time-dependent transformations in Lagrangian}
\textsl{mechanics is converthing it into Einsteinian mechanics}. To do this
conversion, we must: \ a) to replace $\mathcal{L}$ by $u\mathcal{L},$ b) \ to
replace $\dot{q}_{i}$ by $\ q_{i}^{\prime}$ $/u,$ where $q_{i}^{\prime}%
$=$\frac{d}{d\theta}q_{i}$, c) to rewrite new Lagrangian in terms of \ new
time variables and, finally, d) to obtain the Lagrangian equations according
to just described rules, provided that now we have to use $p_{i}^{\prime}$
instead of $\dot{p}_{i}$. In the case if the total energy of the system is
conserved, we shall obtain back the same form of Lagrangian equations
rewritten in terms of new variables. \ From here it follows that by going from
the Lagrangian to Hamiltonian mechanics we can write Hamilton's equations in
which the dotted variables are replaced by the primed. Furthermore, Hamilton's
equations will remain the same even if we replace the Hamiltonian $H$ by some
nonnegative function $f(H)$ while changing time $t$ to time $\theta$ according
to the rule $d\theta/dt=df(H)/dH\mid_{H=E}$. Such a change while leaving
classical mechanics form-invariant will affect quantum mechanics where the
Schr\"{o}dinger's equation%
\begin{equation}
i\hbar\frac{\partial}{\partial t}\Psi=\hat{H}\Psi\tag{45}%
\end{equation}
is \ now replaced by
\begin{equation}
i\hbar\frac{\partial}{\partial\theta}\Psi=f(\hat{H})\Psi. \tag{46}%
\end{equation}
Thus the time changes just described are consistent with both classical and
quantum cases. But the time changes in classical mechanics case are equivalent
to switching from Lagrangian to Einsteinian mechanics. Accordingly, equation
(46) should describe quantum effects at the level of Einsteinian mechanics.
This topic will be discussed further in this section when we \ describe
Bertrand's space-times. With such information at our hands we would like to
discuss the extent to which symmetries of our (empty) space-time affect
dynamics of particles ``living'' in it.

\subsection{Space and Time in Quantum Mechanics}

Use of group-theoretic methods in quantum mechanics was initiated by Pauli in
1926. He obtained complete quantum mechanical solution for\ the Hydrogen atom
employing symmetry arguments only. His efforts were not left without
appreciation. Historically important references can be found in two
comprehensive review papers by Bander and Itzykson [$52$]. In this subsection
we pose and \ formally solve the following problem:

\textsl{Provided that the symmetry of (classical or quantum) system is known,
will this information be sufficient for determination of this system uniquely?
}

Below, we shall provide simple and concrete examples illustrating meaning of
the word ``determination''. In the case of quantum mechanics this problem is
known as the problem about hearing of the ``shape of the drum'' and is
attributed to Mark Kac [$53$].The problem can be formulated as follows.

Suppose that the sound spectrum of the drum is known, will such an information
determine the shape of the drum uniquely? The answer is ``No'' [$54$]. We would
like to explain this non uniqueness using arguments much simpler than those
used by Kac. For this purpose, we choose the most studied example of Hydrogen
atom. The Keplerian motion of a particle (electron) in the centrally symmetric
field is planar \ and is exactly solvable for both the scattering and bound
states at the classical level [$50$]\textbf{. }The result of such a solution
depends on two parameters: the energy and the\textbf{\ }angular\textbf{\ }%
momentum. The correspondence principle formulated by Bohr is expected to
provide the bridge between the classical and quantum realities by requiring
that in the limit of large quantum numbers the results of quantum and
classical calculations for observables should coincide. However, this
requirement may or may not be possible to implement. It is violated already
for the Hydrogen atom! Indeed, according to naive canonical quantization
prescriptions, one should begin with the \textsl{classical} Hamiltonian in
which one has to replace the momenta and coordinates by their operator
analogs. Next, one uses such constructed quantum Hamiltonian in the
Schr\"{o}dinger's equation, etc. Such a procedure breaks down at once for the
Hamiltonian of Hydrogen atom since the intrinsic planarity of the classical
\ Kepler's problem is entirely ignored thus leaving the projection of the
angular momentum without its classical analog. Accordingly, the scattering
differential\textsl{ }crossection for the Hydrogen atom obtained quantum
mechanically\textsl{ }(within the 1st Born\textsl{ }approximation)\textsl{
}uses essentially 3-dimensional calculations in order to obtain the\textsl{
}result\textsl{ }by Rutherford obtained for planar configurations using
classical mechanics! Thus, even for the Hydrogen atom classical and quantum
(or, better, pre quantum) Hamiltonians\textbf{\ }\textsl{do not} match thus
formally violating the correspondence principle. Evidently, semiclassically we
can only think of energy and the angular momentum thus leaving the angular
momentum projection undetermined. Such a ``sacrifice'' is justified by the
agreement between the observed and the predicted Hydrogen atom spectra and by
use of the Hydrogen-like atomic orbitals for multielectron atoms, etc.
Although, to our knowledge, such a mismatch is not mentioned in any textbooks
on quantum mechanics, its existence is essential if we are interested in
extension of these ideas to quantum dynamics of Solar system. In view of such
interest, we would like to reconsider traditional treatments of Hydrogen atom,
this time being guided \ only by the symmetry considerations.

In April of 1940 Jauch and Hill\textbf{\ [}$55$\textbf{]} published a paper in
which they studied the\ planar Kepler problem quantum mechanically. Their work
was stimulated by earlier works by Fock of 1935 and by Bargmann of 1936 in
which it was shown that the spectrum of bound states for the Hydrogen atom can
be obtained by using representation theory of SO(4) group of rigid rotations
of 4-dimensional Euclidean space while the spectrum of scattering states can
be obtained by using the Lorentzian-type group SO(3,1). By adopting results of
Fock and Bargmann to the planar configuration Jauch and Hill obtained the
anticipated result: In the planar case one should use SO(3) group for the
bound states and SO(2,1) group for the scattering states. Although \ this
result will be reconsidered, we mention it now having several purposes in mind.

First, we would like to reverse arguments leading to the final results of
Jauch and Hill in order to return to the problem posed at the beginning of
this section. That is, we want to use the fact that the Kepler problem is
planar (due to central symmetry of the force field) and the fact that the
motion takes place in (locally) Lorentzian space-time in order to argue that
the theory of group representations for Lorentzian SO(2,1) symmetry
group-intrinsic for this type of Kepler problem- \ correctly reproduces the
Jauch-Hill spectrum. Nevertheless, the question remains: Is Kepler's problem
the only one exactly solvable classical and quantum mechanical problem
associated with the SO(2,1) group? Below we argue that this is \textsl{not}
the case! In \ anticipation of such negative result, we would like to develop
our intuition by using some known results from quantum mechanics.

For the sake of space, we consider here only the most generic (for this work)
example in some detail- the radial Schr\"{o}dinger equation for the planar
Kepler problem with the Coulombic potential. It is given by\footnote{The
rationale for discussing the Coulombic potential instead of gravitational will
be fully explained in the next section.}%
\begin{equation}
-\frac{\hbar^{2}}{2\mu}(\frac{d^{2}}{d\rho^{2}}+\frac{1}{\rho}\frac{d}{d\rho
}-\frac{m^{2}}{\rho^{2}})\Psi(\rho)-\frac{Ze^{2}}{\rho}=E\Psi(\rho). \tag{47}%
\end{equation}
Here $\left\vert m\right\vert =0,1,2,...$ is the angular momentum quantum
number as required. For $E<0$ it is convenient to introduce the dimensionless
variable $x$ via $\rho=ax$ and to introduce the new wave function: $\psi
(\rho)=\sqrt{\rho}\Psi(\rho)$. Next, by the appropriate choice of constant $a$
and by redefining $\psi(\rho)$ as $\psi(\rho)=\gamma x^{\frac{1}{2}+\left\vert
m\right\vert }\exp(-y)\varphi(y),$ where $y=\gamma x,$ -$\gamma^{2}=\frac{2\mu
E}{\hbar^{2}}a^{2},a=\frac{\hbar^{2}}{\mu ZE},$ the following hypergeometric
equation can be obtained:%
\begin{equation}
\left\{  y\frac{d^{2}}{dy^{2}}+2[\left\vert m\right\vert +\frac{1}{2}%
-y]\frac{d}{dy}+2[\frac{1}{\gamma}-\left\vert m\right\vert -\frac{1}%
{2}]\right\}  \varphi(y)=0. \tag{48}%
\end{equation}
Formal solution of such an equation is given by $\varphi(y)=\mathcal{F}%
(-A(m),B(m),y),$ where $\mathcal{F}$ is the confluent hypergeometric function.
Physical requirements imposed on this function reduce it to be a polynomial
leading to the spectrum of the planar Kepler problem. Furthermore, by looking
into standard textbooks on quantum mechanics, one can easily find that
\textsl{exactly the same type of hypergeometric equation} is obtained for
problems such as one-dimensional Schr\"{o}dinger's equation with the
Morse-type potential,\footnote{That is, $V(x)=A(exp(-2\alpha x)-2exp(-\alpha
x)).$} three dimensional radial Schr\"{o}dinger equation for the harmonic
oscillator\footnote{That is, $V(r)=\dfrac{A}{r^{2}}+Br^{2}.$} and even three
dimensional radial equation for the Hydrogen atom\footnote{That is,
$V(r)=\dfrac{A}{r^{2}}-\dfrac{B}{r}.$}. Interestingly enough, at the classical
level the last two potentials are the only potentials for which dynamical
trajectories are closed. This is the content of the Bertrand theorem to be
considered in the next subsection.

Since the two-dimensional Kepler problem is solvable with help of
representations of SO(2,1) group, the same should be true for all quantum
problems just listed. That this is the case is demonstrated, for example, in
the book by Wybourne [$56$]. A sketch of the proof is provided in Appendix A.
This proof indicates that, actually, the \textsl{discrete spectrum} of all
problems just listed is obtainable with help of \ representations of SO(2,1)
group! The question \ still remains: If the method discussed in Appendix A
provides the spectra of several quantum mechanical problems listed above, can
we be sure that these are the only exactly solvable quantum mechanical
problems associated with the SO(2,1) group? \ Unfortunately, the answer is
``No''! as we are about to explain.

In Appendix A \ we provide a sketch of the so called spectrum-generating
algebras (SGA) method. It is aimed at producing the exactly solvable
one-variable quantum mechanical problems. In this subsection we would like to
put these results into a broader perspective. In particular, in our \ recent
works [$37,38$] we demonstrated that\textsl{\ all exactly} \textsl{solvable
quantum mechanical problem should involve hypergeometric functions of single
or multiple arguments}. We argued that the difference between different
problems can be understood topologically in view of known relationship between
the hypergeometric functions and braid groups. These results, even though
quite rigorous, are not well adapted for immediate practical use in this
paper. More useful in the present case would be to solve the following problem:

\textsl{For a given} \textsl{set of orthogonal polynomials find the
corresponding many-body operator for which such set of orthogonal polynomials
forms the complete set of} \textsl{eigenfunctions}.

At the level of orthogonal polynomials of one variable relevant for all
exactly solvable two-body problems of quantum mechanics, one can think about
the related problem of finding all potentials in one-dimensional radial
Schr\"{o}dinger's equation, e.g. equation (A.1), leading to the
hypergeometric-type solutions. Very fortunately, such a task was accomplished
already by Natanzon [$57$]. Subsequently, his results were re investigated by
many authors with help of different methods, including SGA. To our knowledge,
the most complete recent summary of the results, including potentials and
spectra can be found in the paper by Levai [$58$]. Even this (very
comprehensive) paper does not cover all aspects of the problem. For instance,
it does not mention the fact that these results had been extended to the
relativistic equations such as Dirac and Klein-Gordon for which similar
analysis was made by Cordero with collaborators [$59$]. In all cited cases
(relativistic and non relativistic) the underlying symmetry group was SO(2,1).
The results of Appendix A as well as of all other listed references can be
traced back to the classically written papers by Bargmann [$60$] and Barut and
Fronsdal [$61$] on representations of SO(2,1) Lorentz group. Furthermore,
subsequently discovered connection \ of this problematics with supersymmetric
quantum mechanics [$62,63$] can be traced back to the 19th century works by
Gaston Darboux. The fact that representations of the \textsl{planar} SO(2,1)
Lorentz group are sufficient to describe all known exactly solvable two-body
problems (instead of the full SO(3,1) Lorentz group!) is remarkable and
intuitively unexpected. It is also sufficient for the purposes of this work
but leaves open the question : Will use of the full Lorentz group produce
exactly solvable quantum mechanical problems not accounted by the SO(2,1)
group symmetry? \ Since the answer to this problem is not affecting the
results of the next section, we leave study of this problem outside the scope
of this work. Instead, we would like to provide an independent arguments in
support of just obtained results.

\subsection{Bertrand Spacetimes and the Issue of Dark Matter }

We begin with \ Einstein equation (33) and \ Newton equation (39). The second
one is obtained from the first by the appropriate change of time variable.
\ The question arises: is it legitimate to make such a comparison? The answer
was in fact already obtained: If we require the Lagrangian equations to stay
form invariant by making the appropriate time changes (not permissible within
the absolute time of Newtonian mechanics) then the distinction between
mechanics of general relativity and Newtonian mechanics becomes blurred.
However, the correspondence between equations (33) and (38) was obtained just
for two body -type problem. The question remains: Can these results be
extended to the many-body case? \ Superficially, it can in the Einstein (that
is massless) limit. This is so because the situation in this limit resembles
Hartree-like\ approximation of quantum mechanics as it was noticed in section
2. Thus, the main question remains: Since the Hartree-type approximation in
quantum mechanics can be systematically improved, could it be that the same is
true with results of general relativity, especially at the scales of Solar
system? Stated alternatively: Can Newtonian dynamics be used at the scales of
our Solar system and general relativity be used \textsl{only} at larger scales
? \ Thus far this is exactly the case in existing physical reality. However,
now we would like to demonstrate that Einsteinian theory of gravity can be
used at any macroscopic space-time scales.

To accomplish this task we need to bring some results from [$64$]. In this
reference the extended phase space was introduced. Following [$33$] and [$65$]
consider the contact 1-form
\begin{equation}
dS=%
{\displaystyle\sum\limits_{i=1}^{n}}
p_{i}dq_{i}-H(p,q)dt. \tag{49}%
\end{equation}
Here $S$ is the classical action while $q_{i}$ and $p_{i}$ are the generalized
coordinates and momenta entering the Hamiltonian $H(p,q)$. For the
conservative system $H(p,q)=E>0.$ We now will treat time $t$ and $E$ (or $H$)
as \ canonically conjugate variables so that $t=q_{n+1}$ and $-E=p_{n+1.}$
Thus, we can write
\begin{equation}
S=%
{\displaystyle\int}
{\displaystyle\sum\limits_{i=1}^{n+1}}
p_{i}dq_{i} \tag{50}%
\end{equation}
Since the Lagrangian $\mathcal{L}$ for such an extended system is given by
\[
\mathcal{L}=\frac{1}{2}%
{\displaystyle\sum\limits_{i,j=1}^{n+1}}
g_{ij}(q)\dot{q}^{i}\dot{q}^{j}-V(q)
\]
where we defined $\dot{q}^{i}=\frac{d}{d\tau}q^{i}$ with the world (or proper)
time $\tau$ (e.g. see (54b)). Next, since
\begin{equation}
p_{i}=\frac{\partial L}{\partial\dot{q}^{i}}=%
{\displaystyle\sum\limits_{k=1}^{n+1}}
g_{ik}(q)\dot{q}^{k} \tag{51}%
\end{equation}
we can write $%
{\displaystyle\sum\limits_{i=1}^{n+1}}
p_{i}dq_{i}=%
{\displaystyle\sum\limits_{i,k}}
g_{ik}(q)\dot{q}^{i}dq^{k}$ and, therefore, we obtain:
\begin{equation}
S=%
{\displaystyle\int}
{\displaystyle\sum\limits_{i=1}^{n+1}}
p_{i}dq_{i}=%
{\displaystyle\int}
{\displaystyle\sum\limits_{i,k}}
g_{ik}(q)\frac{dq^{i}}{d\tau}dq^{k}. \tag{52}%
\end{equation}
In view of the fact that
\begin{equation}
E=\frac{1}{2}%
{\displaystyle\sum\limits_{i,j=1}^{n+1}}
g_{ij}(q)\dot{q}^{i}\dot{q}^{j}+V(q), \tag{53}%
\end{equation}
the result of major importance follows
\begin{equation}
d\tau=\sqrt{\frac{%
{\displaystyle\sum\limits_{i,k}}
g_{ik}(q)dq^{i}dq^{k}}{2(E-V)}}. \tag{54a}%
\end{equation}
Upon substitution of this result into (52) we obtain%
\begin{equation}
S=%
{\displaystyle\int}
\sqrt{2(E-V)%
{\displaystyle\sum\limits_{i,k}}
g_{ik}(q)dq^{i}dq^{k}}\equiv%
{\displaystyle\int}
d\tau\sqrt{%
{\displaystyle\sum\limits_{i,k}}
\tilde{g}_{ik}(q)\dot{q}^{i}\dot{q}^{k}}\equiv%
{\displaystyle\int}
d\rho. \tag{54b}%
\end{equation}
Variation of the action functional $S$ produces equation for geodesics as is
well known [$33,47$]. Since by design, when V=0 the metric tensor $\tilde{g}%
_{ik}$ is having the Lorentzian signature, e.g. see equation (49), in the case
of one test particle the equation for geodesics is describing the motion of
such a particle \ in \ 3+1 dimensional Einsteinian space-time. Furthermore, by
combining (54a) and (54b) we get $d\rho=\sqrt{2(E-V)}d\tau.$ This result is in
qualitative accord with the previously obtained, e.g. $dt=u(\{q_{i}%
\})d\theta.$ \ Evidently, \ for the many-body system \ the motion is taking
place on geodesics in Einsteinian spaces of dimensionality higher than 4.
\ Higher dimensional gravity was considered already in many
places\footnote{e.g. read http://www.phy.olemiss.edu/\symbol{126}%
luca/Topics/grav/higherdim.html}. In the above expressions not only time can
be changed (which is in this extended formalism is part of the configurational
space! ) but also space. This can be seen from the fact that variation of the
action functional $S$ can be written as the geodesic equation
\begin{equation}
\frac{d^{2}q^{k}}{d\tau^{2}}+\Gamma_{ij}^{k}\dot{q}^{i}\dot{q}^{j}=0 \tag{55a}%
\end{equation}
or, as the Lagrangian equation [$47$], page 7,
\begin{equation}
\frac{d}{d\tau}\frac{\partial\mathcal{L}}{\partial\dot{q}^{i}}=\frac
{\partial\mathcal{L}}{\partial q^{j}},\text{ where }\mathcal{L}=\sqrt{%
{\displaystyle\sum\limits_{i,k}}
\tilde{g}_{ik}(q)\dot{q}^{i}\dot{q}^{k}}. \tag{55b}%
\end{equation}
It can be demonstrated [$47$], pages 108-109, that it is possible not to use
the square root for $\mathcal{L}$ for obtaining the same Lagrangian equations.
This fact was effectively used for derivation of the perihelion shift of
Mercury in [$47$], pages 99-108. Thus, the Lagrangian mechanics is
relativistically covariant, as stated already! This gives us the following
protocol of calculations. We illustrate it on example of perihelion of Mercury calculations.

1.Begin with the Landau -Lifshitz \ Lagrangian for a ``free'' particle of unit
mass moving in the plane [$65$]%
\begin{equation}
\mathcal{L}=\frac{1}{2}[(\frac{dr}{dt})^{2}+r^{2}(\frac{d\varphi}{dt})^{2}]
\tag{56a}%
\end{equation}

2. Replace the Newtonian time $t$ by the world time $\tau$ while replacing the
Lagrangian in (56) by the extended Lagrangian (we work in units in which the
speed of light $c=1$)%
\begin{equation}
\mathcal{L}=\frac{1}{2}[(\frac{dr}{d\tau})^{2}+r^{2}(\frac{d\varphi}{d\tau
})^{2}-(\frac{dt}{d\tau})^{2}] \tag{56b}%
\end{equation}

3. Make space and time contact transformations (used in Lagrangian
mechanics[$64$]) with the help of factors e$^{\lambda(r,t)}$ and e$^{\nu
(r,t)}$ \ that is
\begin{equation}
\mathcal{L}=\frac{1}{2}[e^{\lambda(r,t)}(\frac{dr}{d\tau})^{2}+r^{2}%
(\frac{d\varphi}{d\tau})^{2}-e^{\nu(r,t)}(\frac{dt}{d\tau})^{2}] \tag{56c}%
\end{equation}
For the particle in the central field V$\sim\frac{-const}{r}$ the conbination
E-V can always be brought to $1+\frac{const^{\prime}}{r}.$ The presence of
square root cause only a minor change $1+\frac{const^{\prime}}{r}%
\rightarrow1+\frac{1}{2}\frac{const^{\prime}}{r}.$ In which case the rest of
calculations is described, for instance either in [$47$] or in [$22$],
resulting in equation (26) as required. \ The space-time metric which is used
is known as the Schwarzshield metric. We just demonstrated that its origin is
deeply rooted in classical mechanics. At the same time, we just demonstrated
that deep down there is no such a thing like the conventional classical
mechanics, while the extended classical mechanics coincides with Einsteinian.
The question then arises: Is there other uses of classical mechanics which we
overlooked thus far? \ The answer is ``yes''. It is coming from the Bertrand
theorem. Although it is discussed in [$21$], the formulation is a bit vague.
\ We quote this theorem using [$66$].\bigskip

\textbf{Theorem} (Bertrand) \textit{Let }$H=\frac{1}{2}\left\Vert
\mathbf{p}\right\Vert ^{2}+V(\mathbf{q})$\textit{ be a spherically symmetric
Hamiltonian system in the domain of R}$^{3}.$\textit{ Suppose that }

\textit{(i) \ There exist stable circular orbits.}

\textit{(ii) \ All the bounded (other than circular) trajectories are closed.}

\textit{Then the potential }$V(\mathbf{q})$\textit{ is either a Kepler-type:
}$V(\mathbf{q})=A/\left\Vert \mathbf{q}\right\Vert,$\textit{ or a harmonic
oscillator-type }$V(\mathbf{q})=A\left\Vert \mathbf{q}\right\Vert ^{2}.$

Put it simply: in celestial mechanics the closed orbits of two -body problem
with Kepler potential are either circular or ellipsoidal. That is to say,
\textsl{if the potential is strictly Kepler-like then the precession of orbits
cannot occur. \ }In [$8$] it was demonstrated\textsl{ }that for circular
orbits (e.g. read (i) above) the Schwarzshield metrics is compatible with
Kepler's third law. But the Bertrand theorem is responcible for closed
elliptic orbits as well! The Schwarzshield metric is incompatible with
elliptic orbits due to precession. As discussed in section 3, the precession
is result of both a) influence of other planets (which Bertrand's theorem is
not capable to take into account) and b) influence of curvature of space-time
caused by Sun. Thus, at the level of planetary motion in our Solar sytem
Bertrand's theorem serves only as some reasonable approximation.
\ Improvements of Bertrand's theorem had been recently discussed [$67$]. They
lead to the differential equations of Hill-type. These are discussed further
in section 5.4. in connection with rings of heavy planets. Interestingly
enough the motion of stars around the massive galaxy center is taking place on
circular orbits according to recent astronomical observations [$68$]. The
question then emerges: Is general relativity capable of describing such an
observed motion of stars? Put it differently: \ \textsl{Is there space-times
consistent with formalism of general relativity in which the geodesic motion
can take place on circular orbits? }The answer was provided in the seminal
paper by Perlic [$69$] done in 1992. He was the first to call these
space-times as \textsl{Bertrand spacetimes.} In his analysis\textsl{ }of the
obtained results Perlic stated that \ although ``relativity allows of a
hypotetical body that produces (if isolated from other sourses) a field in
which all bounded trajectories are periodic,\textsl{ but this body must be a
rather exotic object.}'' Subsequent studies (both observational and
theoretical) revealed that this exotic object is nothing else but the dark
matter. \textsl{Thus the dark matter can be described within Einstenian
general relativity}! Because of this, it is worth to describe briefly the observational and
theoretical studies. They will serve us well in the rest of this paper.

The major observational discovery leading to the concept of dark matter
is as follows. \ Most of astronomers used Newtonian
mechanics at the scales of individual galaxies and even beyond. To interpret
the observational data the model was needed based on laws of Netonian
mechanics. Initially the galaxy was modelled on example of our Solar system in
which individual planets are moving on Keplerian orbits. The quantity of major
interest is \textsl{the rotation curve} plot. It is made in coordinates such
that x-axis represents the distance $r$ from the galaxy center while the
y-axis represents the observed velocity $v$ of, say, some individual star. If
one believes Newton, then simple calculation yields $v\sim\frac{1}{r}$ type
plot. \ This is NOT what was observed in the sky though! Next came other
models, e.g. the rigid body model. If the galaxy can be modelled by a pancake,
then $v\sim r.$ \ Observations demonstrate that this is also NOT the case!
\ What is observed in the sky is the following. At space scales not too
distant from the galaxy center the law $v\sim r$ is  observed. At some
threshold distance from the center, this behavior is smoothly crossovering
into $v\sim const$ \ Such a dependence holds for scales almost up to the edges
of galaxy and only at the very edge of galaxy one observes the Newtonian
behavior. \ Rotation curves happen to be of major importance since they allow
estimation of the galactic masses. To explain the observed behavior of
rotation curves the suggestion came that each galaxy is surrounded outside
(and, in part, penetrated inside) by the dark matter, that is by the
substance\ invisible optically from the Earth. The situation here resembles
very much that we considered when discussed Leverier-type calculations in
section 3.1. \ Being armed with Perlic's results, the authors of [$68$] obtained
\ the rotation curve which is in very good agreement with the observational data.
Their work is one of many of various degree of sophistication which will be
discussed elsewhere. What is of interest in this work is the fact that the
rotation curve was obtained based on Perlick's results. Furthermore, Perlick's
work begins with the Lagrangian given by the equation (56c)! \ Since, as we
already demonstrated, this Lagrangian is very general, it admits further
general study, just like that discussed in standard classical mechanics
textbooks. Specifically, using standard mechanical calculations the above
Lagrangian for the ``free particle'' is also an energy
\begin{equation}
E=\frac{1}{2}[e^{\lambda(r,t)}\dot{r}^{2}+r^{2}\dot{\varphi}^{2}-e^{\nu
(r,t)}\dot{t}^{2}] \tag{57}%
\end{equation}
In addition, the angular momentum $M$ is conserved. That is
\begin{equation}
M=r^{2}\dot{\varphi}. \tag{58}%
\end{equation}
Now, however, not only M is cyclical variable but also $C$ defined by
\begin{equation}
C=e^{\nu(r,t)}\dot{t} \tag{59}%
\end{equation}
Use of natural (proper) parametrization leads to the condition
\begin{equation}
C=1. \tag{60}%
\end{equation}
Since equations (56c)-(60) are eaxactly the same as those used by Perlic, there
is no need to duplicate the rest of his mechanical analysis here. Naturally,
this analysis is aimed at obtaining $\lambda(r,t)$ and $\nu(r,t)$ explicitly
\ and, by doing so, at obtaining of the Einsteinian metric. Two types of metric had been
obtained. For description of circular motion of stars around the galaxy center
[$68$] the already familiar (to us) metric was obtained. In notations of
[$68$] it is given by (planar version)
\begin{equation}
ds_{BST}^{2}=-\frac{dt^{2}}{D+\frac{\alpha}{r}}+\frac{dr^{2}}{\beta^{2}}%
+r^{2}d\varphi^{2}. \tag{61}%
\end{equation}
Here $D$,$\alpha$ and $\beta$ are parameters (constants) and BST stands for
''Bertrand space-time''. \ From the previous discussion we recognize the factor
$D+\frac{\alpha}{r}$ as E-V. \ By looking at equation (53) we recognize that
in the nonrelativistic case we would have a relationship
\[
(\frac{dr}{dt})^{2}+r^{2}(\frac{d\varphi}{dt})^{2}-\frac{\alpha}{r}=E
\]
or
\begin{equation}
dr^{2}+r^{2}d\varphi^{2}-(E+\frac{\alpha}{r})dt^{2}=0. \tag{62a}%
\end{equation}
In the extended configurational space we obtain instead%
\begin{equation}
dt^{2}=dr^{2}+r^{2}d\varphi^{2}-(E+\frac{\alpha}{r})d\tau^{2} \tag{62b}%
\end{equation}
By looking at (57) and (60) we can bring equation (62b) into form coinciding
with that given in (61) (for $\beta=1).$ Finally, it is permissible to replace
$dr^{2}$ by $\frac{dr^{2}}{\beta^{2}}$ in view of (57). Physically, such a
replacement leads to introduction of some characteristic spatial scale. Now we
are ready to proceed with our next task-quantization of Solar System dynamics.

\section{Some Classical and Quantum Aspects  of Solar System\\ Dynamics}

\subsection{General Comments}

Attentive reader probably already noticed that at the heart of quantization
lies the honeycomb condition, Eq.(4) (or (7)), allowing to restore both
Heisenberg's and Schrodinger's versions of quantum mechanics. In the case of
celestial mechanics the honeycomb condition \ can be equivalently rephrased as
the condition of \textsl{planetary alignment} [$8$].This boils down to the study of
the issue of possibility of some or all planets to be aligned on the same line.
Very much like with dark matter, this issue cannot be resolved purely
theoretically. Experimentally, this issue is resolved positively to a large
extent (that is to the extent with which measurements were made) thus leading to
relations of the type given by equation (4). The same is true for the
satellites of heavy planets. In section\ 4.2. we demonstrated that the
quantum-classical correspondence could be sometimes not one-to-one. \ In such
cases, to reduce the number of options some additional physical arguments
should be used. We shall provide these arguments in this section. We begin by
noticing that while Eq.(4) (or (7)) is only sufficient condition for
quantization, the necessary condition in atomic and celestial mechanics lies
in the \textsl{non dissipativity}. Indeed, thus far we developed our formalism
only for systems in which energy is not conserved. Recall that Bohr introduced
his quantization prescription to avoid dissipation caused by the emission of
radiation \ by electrons in orbits in general position. New quantum mechanics
have \textsl{not} \ shed much light on the absence of dissipation for
stationary Bohr's orbits. At the level of old Bohr theory absence of
dissipation at the stationary Bohr orbit was explained by Boyer [$70$].
Subsequently, his result was refined by Puthoff\textbf{\ [}$71$\textbf{]}. In
the case of Solar system absence of dissipation for motion on stable orbits
(geodesics) was discussed by Goldreich [$72$] who conjectured that the
\ dissipative (tidal) effects adjust the initial motion of planets/satellites
in such a way that eventually the orbits become stable. Notice that dynamics
of Solar System\ as considered by Poincare$^{^{\prime}}$ and by those who
developed his ideas does not involve treatment of tidal effects. Treatment of
tidal effects in general relativity is discussed in [$10$] and requires
consideration of motion of extended objects, e.g. read Papapetrou, Ref.
[$46$]. In view of results of Section 3, following Einstein, we shall ignore
finite sizes of planets and/or satellites of heavy planets thus removing the
problem of tidal friction. This is justified by the numerical results to be
obtained in this section.

Thus, following Bohr, \textsl{we postulate} that in the case of Solar System
dynamics on stable orbits is non dissipative. This assumption then leads us to
the following Table 1. In this table by accidental degeneracy we mean the
condition given by Eq.(4) which is the same as the planetary alignment
condition\footnote{E.g. read
http://www.faqssys.info/could-this-planetary-superalignment-happen/}.\

\begin{table}[ht]%
\caption{}
{\footnotesize{\begin{tabular}
[t]{|l|l|l|}\hline%
\begin{tabular}
[c]{l}%
$\backslash$%
$Type$\textit{\ }$of$\textit{\ }$mechanics$\\
$\mathit{Properties}$%
\end{tabular}
&
\begin{tabular}
[c]{l}%
$Quantum$\textit{\ }$atomic$\\
$mechanics$%
\end{tabular}
&
\begin{tabular}
[c]{l}%
$\mathit{Quantum}$\\
$celestial$\textit{\ }$mechanics$%
\end{tabular}
\\\hline%
\begin{tabular}
[c]{l}%
Dissipation (type of)%
$\backslash$%
\\
(yes%
$\backslash$%
no)%
$\backslash$%
on stable orbits
\end{tabular}
&
\begin{tabular}
[c]{l}%
electromagnetic\\
friction%
$\backslash$%
no%
$\backslash$%
\\
Bohr orbits
\end{tabular}
&
\begin{tabular}
[c]{l}%
gravitational radiation\\%
$\backslash$%
no%
$\backslash$%
Einstein's geodesics
\end{tabular}
\\\hline%
\begin{tabular}
[c]{l}%
Accidental degeneracy%
$\backslash$%
\\
(yes%
$\backslash$%
no)%
$\backslash$%
origin
\end{tabular}
& yes%
$\backslash$%
Bohr-Sommerfeld condition & yes%
$\backslash$%
Laplace-Lagrange condition\\\hline
Charge neutrality & yes & no(but see below)\\\hline
Masses &
\begin{tabular}
[c]{l}%
electrons are having\\
the same masses
\end{tabular}
&
\begin{tabular}
[c]{l}%
(up to validity of the\\
equivalence principle)\\
masses are the same
\end{tabular}
\\\hline
Minimal symmetry group & SO(2,1) & SO(2,1)\\\hline
Correspondence principle & occasionally violated & occasionally
violated\\\hline%
\begin{tabular}
[c]{l}%
Discrete spectrum:\\
finite or infinite%
$\backslash$%
reason%
$\backslash$%
\\
Pauli principle(yes%
$\backslash$%
no)
\end{tabular}
&
\begin{tabular}
[c]{l}%
finite and infinite%
$\backslash$%
\\
charge neutrality%
$\backslash$%
\\
yes
\end{tabular}
&
\begin{tabular}
[c]{l}%
finite%
$\backslash$%
\\
no charge neutrality%
$\backslash$%
\\
yes
\end{tabular}
\\\hline
\end{tabular}}}
\end{table}

\subsection{Celestial Spectroscopy and the 
Titius-Bode Law of Planetary Distances}

The atomic spectroscopy was inaugurated by Newton in the second half of 17th
century. The celestial spectroscopy was inaugurated by Titius in the second
half of 18th century and become \ more famous after it was advertised by
Johann Bode, the Editor of the ``Berlin Astronomical Year-book''. The book by
Nieto [$73$] provides extensive bibliography related to uses and
interpretations of the Titius-Bode (T-B) law up to second half of 20th
century. Unlike the atomic spectroscopy, where the observed atomic and
molecular spectra were expressed using simple empirical formulas which were
(to our knowledge) never elevated to the status of \ ``law'', in celestial
mechanics the empirical T-B formula
\begin{equation}
r_{n}=0.4+03.\cdot2^{n}\text{, \ \ \ }n=-\infty,0,1,2,3,... \tag{63}%
\end{equation}
for the orbital radii (semimajor axes) of planets acquired the status of a law
in the following sense. In the case of atomic spectroscopy the empirical
formulas used for description of atomic/molecular spectra have not been used
(to our knowledge) for making predictions. Their purpose was just to describe
in mathematical terms what had been already observed. Since the T-B empirical
formula for planetary distances was used as the law, it was used in search for
planets not yet discovered. In such a way Ceres, Uranus, Neptune and Pluto
were found [$74$]. However, the discrepancies for Neptune and Pluto were much
larger than the error margins allowed by the T-B law. This fact divided the
astronomical community. Without going into historical details, we would like
to jump to the very end of the Titius-Bode story in order to use its latest
version \ which we found in the paper by Neslu\v{s}an [$75$] who, in turn, was
motivated by the work of Lynch [$76$]. Instead of (63) these authors use
another empirical power law dependence
\begin{equation}
r_{n}=r_{0}B^{n},\text{ }n=1,2,3,..,9. \tag{64}%
\end{equation}
For planets (except Pluto and including the asteroid belt) Neslu\v{s}an
obtained\footnote{In astronomical units (to be defined below).} $r_{0}%
(au)=0.203$ and $B=1.773$ with the rms deviation accuracy of
0.0534\footnote{This result gives for the Earth in astronomical (au) units the
result $r_{3}\simeq1.13.$ Much better result is obtained in case if we choose
$B=1.7.$ In this case we obtain: $r_{3}\simeq.997339.$ Lynch provides
$B=1.706$ and $r_{0}=0.2139.$}. Analogous power law dependencies were obtained
previously in the work by Dermott [$77$] for both planets and satellites of
heavy planets such as Jupiter, Saturn and Uranus. It should be noted that
\ because of noticed discrepancies the attempts were made to prove or disprove
the Titius-Bode law by using statistical analysis, e.g. see papers by Lynch
[$76$] and Hayes and Tremaine [$78$], with purpose of finding out to which
extent the observed dependencies can be considered as non accidental.
Following \ the logic of Bohr, we would like to use the observed empirical
radial dependencies as a guide for our calculations.

\subsection{An Attempt at   Quantization of Solar System Dynamics}

Being guided by the Table 1 we shall assume that planets do not interact since
they move along geodesics as discussed in Sections 3 and 4. In the case of
atomic mechanics it was clear from the beginning that such an approximation
should sooner or later fail. The nonexisting electroneutrality in the sky
\ provides strong hint that the T-B law must be of very limited use since the
number of discrete levels \ for gravitating systems should be \textsl{always
finite}. Otherwise, we would observe the countable infinity of satellites
around \ Sun or of any of heavy planets. This is \ not observed and therefore
is physically incorrect. It is incorrect because such a system would tend to
capture all matter in the Universe. The same applies to the photon sphere
discussed in Section 3. In it we noticed that classically this orbit (sphere)
is unstable. If such a sphere would be stable quantum mechanically, this would
cause the black hole to grow in mass indefinitely since it would accumulate
incoming photons with the wavelengths lesser than 6$\pi M$ if conditions for
capture are right. \ Whether or not such accumulation is actually possible now
can be investigated in laboratory conditions as discussed already.

In the literature one can find many attempts at quantization of Solar system
using standard prescriptions of quantum mechanics.\ Many of these papers are
listed in [$14$]. However, we do not provide references to papers whose
results do not affect ours. Blind uses of standard rules of quantum mechanics
\ for quantization of Solar system\ dynamics do not contain any provisions
\ for finite number\textsl{ }of\textsl{ }energy levels/orbits for gravitating
systems be it our Solar sytem or some galaxy\textsl{. }To facilitate matters,
in the present case we would like to make several additional observations.
First, we have to find an analog of the Planck constant. Second, we have to
have some mechanical model in mind to make our search for physically plausible
answer successful. To accomplish the first task, we have to take into account
the 3-rd Kepler's law. In accord with Eq.(32), it can be written
as\footnote{We have included the gravitation constant $\gamma$ in this
expression.} $r_{n}^{3}/T_{n}^{2}=\dfrac{4\pi^{2}}{\gamma(M_{\odot}+m)}$. \ In
view of arguments presented in Section 3, we can safely approximate the
r.h.s. by $4\pi^{2}/\gamma M_{\odot}$, where $M_{\odot}$ is the mass of the
Sun. For the \ purposes of this work, it is convenient to restate this law as%
\begin{equation}
3lnr_{n}-2\ln T_{n}=\ln4\pi^{2}/\gamma M_{\odot}=const \tag{65}%
\end{equation}
Below, we choose the \textsl{astronomical system of units} in which $4\pi
^{2}/\gamma M_{\odot}=1.$ By definition, in such system of units we have for
the Earth: $r_{3}=T_{3}=1$. Consider now the Bohr result, Eq.(6), and take
into account that $E=\hbar\omega\equiv\dfrac{h}{2\pi}\dfrac{2\pi}{T}.$
Therefore, Bohr's result can be conveniently restated as $\omega
(n,m)=\omega(n)-\omega(m).$ Taking into account \ Eq.s (6),(46),(64) and the
third Kepler's law, Eq.65), we formally obtain:%
\begin{equation}
\omega(n,m)=\frac{1}{c\ln\tilde{A}}(nc\ln\tilde{A}-mc\ln\tilde{A}), \tag{66}%
\end{equation}
where the role of Planck's constant is being played now by $c\ln\tilde{A}$,
$\tilde{A}=B^{\frac{3}{2}}$ and $c$ is some constant to be determined
selfconsistently below\footnote{Not to be confused with the speed of light
!}$.$

At first, one may think that what we obtained is just a simple harmonic
oscillator spectrum. After all, this should come as not too big a surprise
since both, the Newtonian Eq.(15) and the Einsteinian Eq.(26) (when brought to
the Newtonian-looking form) are classical equations for the harmonic
oscillator (evidently, this is still the content of Bertrand theorem). This
result is also compatible with that of Appendix A. The harmonic oscillator
option is physically undesirable\ though (when one is contemplating
quantization) since the harmonic oscillator has countable infinity of energy
levels. Evidently, such a spectrum is equivalent to the T-B law. But it is
well known that this law is not working for large numbers!

To make progress, we have to use the 3rd Kepler's law once again. This time,
we have to take into account that in astronomical system of units $3lnr_{n}=$
$2\ln T_{n}.$ A quick look at equations (A.11), (A.12) suggests that the
underlying mechanical system is likely to be associated with that for the
Morse potential. \ This is so because the low lying states of such a system
cannot be distinguished from those for the harmonic oscillator. However, this
system does have only a finite number of energy levels which makes sense
physically. The task remains to connect this system with the planar Kepler's
problem. Although in view of results of Appendix A such a connection does
exist, we want to demonstrate it explicitly at the level of classical
mechanics first.

Following Pars [$50$], the motion of a point of unit mass in the field of
\ Newtonian gravity (Kepler potential) is described by
\begin{equation}
\dot{r}^{2}=(2Er^{2}+2\gamma Mr-\alpha^{2})/r^{2}, \tag{67}%
\end{equation}
where $\alpha$ is the angular momentum \footnote{To comply with notations of
the \ book by Pars we replaced $L$ by $\alpha.$} (e.g. see equation (5.2.55)
of [$50$]). Next, we replace $r(t)$ by $r(\theta)$ in such a way that
$dt=u(r$($\theta)$)$d\theta$ \footnote{Notice that time change transforms
these classical results into that of general relativity!}. Let therefore
$r(\theta)=r_{0}\exp(x(\theta)),$ -$\infty<x<\infty.$ Unless otherwise
specified, we shall write $r_{0}=1$. In such (astronomical) system of units)
we obtain, $\dot{r}=x^{\prime}\dfrac{d\theta}{dt}\exp(x(\theta)).$ This result
can be further simplified by choosing $\dfrac{d\theta}{dt}=\exp(-x(\theta)).$
With this choice (67) acquires the following form:%
\begin{equation}
(x^{\prime})^{2}=2E+2\gamma M\exp(-x)-\alpha^{2}\exp(-2x). \tag{68}%
\end{equation}
Consider points of equilibria for the potential \ $U(r)=-2\gamma
Mr^{-1}+\alpha^{2}r^{-2}.$ Using it, we obtain: $r^{\ast}=\dfrac{\alpha^{2}%
}{\gamma M}.$ According to\ $[21$] such defined $r^{\ast}$ coincides with the
major elliptic semiaxis. It can be also shown, e.g. look into [$50$], equation
(5.4.14), that for the Kepler problem the following relation holds:
$\ E=-\dfrac{\gamma M}{2r^{\ast}}$. Accordingly, $r^{\ast}=-\dfrac{\gamma
M}{2E},$ and, furthermore, using the condition $\frac{dU}{dr}=0,$ we obtain:
$\dfrac{\alpha^{2}}{\gamma M}=-\dfrac{\gamma M}{2E}$ or, $\alpha^{2}$
$=-\dfrac{\left(  \gamma M\right)  ^{2}}{2E}.$ Since in the chosen system of
units $r(\theta)=\exp(x(\theta)),$ we obtain as well: $\dfrac{\alpha^{2}%
}{\gamma M}=\exp(x^{\ast}(\theta)).$ It is convenient to choose $x^{\ast
}(\theta)=0.$ This requirement sends the point $x^{\ast}(\theta)=0$ to the
origin of new coordinate system and implies that with respect to such chosen
origin $\alpha^{2}=\gamma M.$ In doing so some caution should be exercised
since, upon quantization, the equation $r^{\ast}=\dfrac{\alpha^{2}}{\gamma M}$
becomes $r_{n}^{\ast}=\dfrac{\alpha_{n}^{2}}{\gamma M}.$ By selecting the
astronomical scale $r_{3}^{\ast}=1$ as the unit of length, implies then that we can
write the angular momentum $\alpha_{n}^{2}$ as $\varkappa$ $\dfrac{r_{n}%
^{\ast}}{r_{3}^{\ast}}$ and \ to define $\varkappa$ as $\alpha_{3}^{2}$
$\equiv\alpha^{2}.$ Using this fact \ Eq.(68) can then be conveniently
rewritten as
\begin{equation}
\frac{1}{2}(x^{\prime})^{2}-\gamma M(\exp(-x)-\frac{1}{2}\exp(-2x))=E
\tag{69a}%
\end{equation}
or, equivalently, as
\begin{equation}
\frac{p^{2}}{2}+A(\exp(-2x)-2\exp(-x))=E, \tag{69b}%
\end{equation}
where $A=\dfrac{\gamma M}{2}.$ \ \ Since this result is exact classical analog
of the quantum Morse potential problem, the transition to quantum mechanics
now can be done straightforwardly. By doing so \ we have to replace the
Planck's constant $\hbar$ by $c\ln\tilde{A}$ according  to Eq.(66). After that, we can write the
answer for the spectrum at once [$79$]:%
\begin{equation}
-\tilde{E}_{n}=\frac{\gamma M}{2}[1-\frac{c\ln\tilde{A}}{\sqrt{\gamma M}%
}(n+\frac{1}{2})]^{2}. \tag{70}%
\end{equation}
This result contains an unknown parameter $c$ to be determined now. To do so,
it is sufficient to expand the potential in \ Eq.(69b) and to keep terms up to
quadratic. Such a procedure produces the anticipated harmonic oscillator
result%
\begin{equation}
\frac{p^{2}}{2}+Ax^{2}=\tilde{E} \tag{71}%
\end{equation}
whose quantum spectrum is given by $\tilde{E}_{n}=(n+\frac{1}{2})c\sqrt{2A}%
\ln\tilde{A}.$ In the astronomical system of units the spectrum reads:
$\tilde{E}_{n}=(n+\frac{1}{2})c2\pi\ln\tilde{A}$ . This result is in agreement
with (66). To proceed, we \ notice that in (66) the actual sign of the
Planck-type constant is undetermined. Specifically, in our case (up to a
constant) the energy $\tilde{E}_{n}$ is determined by ln$\left(  \frac
{1}{T_{n}}\right)  $ $=-\ln$ $\tilde{A}$ so that it makes sense to write
$-\tilde{E}_{n}\sim n\ln\tilde{A}.$ To relate the classical energy defined by the
Kepler-type equation $E=-\dfrac{\gamma M}{2r^{\ast}}$ to the energy we just
have defined, we have to replace the Kepler-type equation by $-\tilde{E}%
_{n}\equiv-\ln\left\vert E\right\vert =-2\ln\sqrt{2}\pi+\ln r_{n}\text{ .\ }%
$This should be done in view of the 3rd Kepler's law and the fact that the new
coordinate $x$ is related to the old coordinate $r$ via $r=e^{x}$. Using
\ Eq.(64) (for $n=1$) in the previous equation and comparing it with the
already obtained spectrum of the harmonic oscillator we obtain:
\begin{equation}
-2\ln\sqrt{2}\pi+\ln r_{0}B=-c2\pi\ln\tilde{A}, \tag{72}%
\end{equation}
where in arriving at this result we have subtracted the nonphysical ground
state energy. Thus, we obtain:%
\begin{equation}
c=\frac{1}{2\pi\ln\tilde{A}}\ln\frac{2\pi^{2}}{r_{0}B}. \tag{73}%
\end{equation}
Substitution of this result back into Eq.(70) produces\footnote{See Eq.(64)
for the numerical data.}%
\begin{align}
-\tilde{E}_{n}  &  =2\pi^{2}[1-\frac{(n+\frac{1}{2})}{4\pi^{2}}\ln\left(
\frac{2\pi^{2}}{r_{0}B}\right)  ]^{2}\simeq2\pi^{2}[1-\frac{1}{9.87}%
(n+\frac{1}{2})]^{2}\nonumber\\
&  \simeq2\pi^{2}-4(n+\frac{1}{2})+0.2(n+\frac{1}{2})^{2}. \tag{74}%
\end{align}
To determine the number of bound states, we follow the same procedure as was
developed long time ago in chemistry for the Morse potential. \ For this
purpose\footnote{Please, take a note that in chemistry the Morse potential is
being routinely used (since early days of quantum mechanics) for description
of vibrational spectra of diatomic molecules.} we introduce the energy
difference $\Delta\tilde{E}_{n}=$ $\tilde{E}_{n+1}-\tilde{E}_{n}=4-0.4(n+1)$
first. Next, the maximum number of bound states is determined by requiring
$\Delta\tilde{E}_{n}=0.$ In our case, we obtain: $n_{\max}=9$. This number is
in perfect accord with observable data for planets of our Solar system (with
Pluto being excluded and the asteroid belt included). In spite of such a good
accord, some caution must be still exercised while analyzing the obtained
result. Should we not insist on physical grounds that the discrete spectrum
must contain only finite number of levels, the obtained spectrum for the
harmonic oscillator would be sufficient (that is to say, that the validity of
the T-B law would be confirmed). Formally, such a choice \ also solves the
quantization problem completely and even is in accord with the numerical data
[$75$]. The problem lies however in the fact that these data were fitted to
the power law, Eq.(64), in accord with the original T-B empirical guess.
Heisenberg's honeycomb rule, Eq. (7b), does \textsl{not} rely on specific
$n-$dependence. \ In fact, we have to consider the observed (the
Titius-Bode-type) $n-$dependence only as a hint, especially because in this
work we intentionally avoid use of any adjustable parameters. The developed
procedure, when supplied with correctly interpreted numerical data, is
sufficient for obtaining results without any adjustable parameters as we just
demonstrated. In turn, this allows us to replace the T-B law in which the
power $n$ is unrestricted by physically more accurate result working
especially well for larger values of $n$. \ For instance, the constant $c$ was
determined using the harmonic approximation for the Morse-type potential. This
approximation is expected to fail very quickly as the following arguments
indicate. Although $r_{n}^{\prime}s$ can calculated using the T-B law given by
Eq.(64), the arguments following this equation cause us to look also at the
equation $-\tilde{E}_{n}\equiv-\ln\left\vert E\right\vert =-2\ln\sqrt{2}%
\pi+\ln r_{n\text{ \ }}$ for this purpose. This means that we have to use
Eq.(74) (with ground state energy subtracted) in this equation in order to
obtain the result for $r_{n}.$ If we ignore the quadratic correction in
Eq.(74) (which is equivalent to calculating the constant $c$ using harmonic
oscillator approximation to the Morse potential) then, by construction, we
recover the T-B result, Eq.(64). If, however, we do not resort to such an
approximation, calculations will become much more elaborate. The final result
will indeed replace the T-B law but the obtained analytical form is going to
be too cumbersome for practitioners. Since corrections to the harmonic
oscillator potential in the case of Morse potential are typically small, they
do not change things qualitatively. Hence, we do not account for these
complications in our paper. Nevertheless, accounting for these (anharmonic)
corrections readily explains why the empirical T-B law works well for small
n's and becomes increasingly unreliable for larger n's [$73,74$].

In support of our way of \ doing quantum calculations, we would like to
discuss now similar calculations for satellite systems of Jupiter, Saturn,
Uranus and Neptune. To do such calculations the astronomical system of units
is not immediately useful since in the case of heavy planets one cannot use
the relation $4\pi^{2}/\gamma M_{\odot}=1.$ This is so because we have to
replace the mass of the Sun $M_{\odot}$ by the mass of respective heavy
planet. For this purpose we write $4\pi^{2}=$\ $\gamma M_{\odot}$ then
multiply both \ sides by $M_{j\text{ }}$\ (where $j$ stands for the $j$-th
heavy planet) and divide both sides by $M_{\odot}$. Thus, we obtain: $4\pi
^{2}q_{j}=$\ $\gamma M_{j}$,\ where $q_{j}=$\ $\dfrac{M_{j}}{M_{\odot}}$\ .
Since the number $q_{j}$ is of order $10^{-3}$\ $-10^{-5}$, it causes some
inconveniences in actual calculations. To avoid this difficulty, we need to
readjust Eq.(69a)\ by rescaling $x$ coordinate as $x=\delta\bar{x}$\ and, by
choosing \ $\delta^{2}$\ $=q_{j}$. After transition to quantum mechanics such
a rescaling results in replacing \ Eq.(70) for the spectrum by the following
result:%
\begin{equation}
-\tilde{E}_{n}=\frac{\gamma M}{2}[1-\frac{c\delta\ln\tilde{A}}{\sqrt{\gamma
M}}(n+\frac{1}{2})]^{2}. \tag{75}%
\end{equation}
Since the constant $c$\ is initially undetermined, we can replace it by
$\tilde{c}=c\delta.$ This replacement \ allows us to reobtain back equation
almost identical to Eq.(74). That is
\begin{equation}
-\tilde{E}_{n}=2\pi^{2}[1-\frac{(n+\frac{1}{2})}{4\pi^{2}}\ln\left(
\frac{\gamma M_{j}}{(r_{j})_{1}}\right)  ]^{2}. \tag{76}%
\end{equation}
In this equation \ $\gamma M_{j}=$\ $4\pi^{2}q_{j}$ and \ $(r_{j})_{1}$ is the
semimajor axis of the satellite lying in the equatorial plane and closest to
the $j$-th planet. Our calculations are summarized in the Table 2 below.
Appendix B contains the input data used in calculations of n$_{theory}^{\ast
}.$\ Observational data are taken from the web link:
\ http://nssdc.gsfc.nasa.gov/planetary/ \ \ Then, go to the respective planet
and, then-to the ``fact sheet'' link for this planet.\ \ \ 

\begin{table}[ht]%
\caption{}
\begin{center}
\begin{tabular}[c]{|l|l|l|}\hline
Satellite system%
$\backslash$%
n$_{\max}$ & n$_{theory}^{\ast}$ & n$_{\text{obs}}^{\ast}$\\\hline
Solar system & 9 & 9\\\hline
Jupiter system & 11-12 & 13\\\hline
Saturn system & 20 & 20\\\hline
Uranus system & 40 & 27\\\hline
Neptune system & 33 & 14\\\hline
\end{tabular}
\end{center}
\end{table}
 
The data are taken  in 2015. Since the discrepancies for Uranus and Neptune systems may be genuine or may be not, we come up with the following general filling pattern to be considered
in the next subsection. In the meantime, we would like to mention that Pluto also has 5 satellites rotating more or less in the same plane. The motion is of resonance nature, e.g. see Eq.(4), inclicating quantum nature of Pluto satellite system. It would be interesting to check this number 5 by using  Eq.(76).

\subsection{Filling patterns in Solar System: Similarities and Differences  with Atomic Mechanics}

From atomic mechanics we know that the approximation of independent electrons
used by Bohr fails rather quickly with increased number of electrons. \ For
this reason alone to expect that the T-B law is going to hold for satellites
of heavy planets is naive. At the same time, for planets rotating around the
Sun such an approximation is seemingly good but also not without flaws. The
SO(2,1) symmetry explains why motion of all planets should be planar but it
does not explain why motion of all planets is taking place in the plane
coinciding with the equatorial plane of the Sun or why all planets are moving
in the same direction. The same is true for the regular satellites of all
heavy planets as discussed by Dermott [$77$]. Such a configuration can be
explained by a plausible hypothesis [80] that all planets of Solar System and
regular satellites of heavy planes are originated from evolution of the
pancake-like cloud. The assumption \ that all planets lie in the same ( Sun's
equatorial) plane and move in the same direction coinciding with the direction
of rotation of the Sun around its axis was \ used essentially by
Poincare$^{\prime}$ in his ``Les Methodes Nouvelles de la Mecanique Celeste''
written \ between 1892 and 1899 [$26$]. These assumptions \ are compatible
with the hypothesis [80] of how Solar system was formed. In 1898, while
Poincare was \ still working on his \ ``Methodes'' the shocking \ counter
example to the Poincare$^{\prime}$ theory was announced by Pickering who
discovered the ninth moon of Saturn (eventually named Phoebe) rotating in the
direction opposite to all other satellites of Saturn. \ Since that time the
satellites rotating in the ``normal'' direction are called ``\textit{regular}''
(or ``\textit{prograde}'') while those rotating in the opposite direction called
``\textit{irregular}''(or ``\textit{retrograde}''). At the time of writing of this
paper 113 irregular satellites were discovered (out of those, 93 were
discovered after 1997 thanks to space exploration by rockets)\footnote{E.g.
read ``Irregular moon'' in Wikipedia}. Furthermore, in the late 2009 \ Phoebe
had brought yet another surprise to astronomers. Two articles in Nature
[81,82] are describing the largest new ring of Saturn. This new ring lies in
the same plane as Phoebe's orbit and, in fact, the \ Phoebe's trajectory is
located inside the ring. The same arrangement is true for regular satellites
and the associated with them rings.

To repair the existing theory of formation of Solar system one has to make an
assumption that all irregular satellites are ``strangers''. That is that they
were captured by the already existing \ and fully developed Solar system. Such
an explanation \ would make perfect sense should the orbits of these strangers
be arranged in a completely arbitrary fashion. But they are not!
\textsl{Without} \textsl{an exception}, it is known that: a) all retrograde
satellite orbits are lying \textsl{strictly outside} of the orbits of prograde
satellites, b) the inclinations of their orbits is noticeably different from
\ those for prograde satellites, however, c) by analogy with prograde
satellites they tend to group (with very few exceptions) in orbits-all having
the same inclination so that different groups of retrogrades are having
differently inclined orbits in such a way that these orbits \ do not overlap
if the retrograde plane of \ satellites with one inclination is superimposed
with that for another inclination\footnote{As before, go to
http://nssdc.gsfc.nasa.gov/planetary/ \ \ \ then, go to the respective planet,
and then-to the ``fact sheet'' link for this planet.}. In addition, all objects
lying outside the sphere made by the rotating plane in which all planets lie
are arranged in a similar fashion[$74$]. Furthermore, the orbits of prograde
satellites of all heavy planets \ lie in the respective equatorial planes-
just like the Sun and the planets - thus forming miniature Solar-like systems.
These equatorial planes are tilted with respect to the Solar equatorial plane
since all axes of rotation of heavy planets are tilted\footnote{That is the
respective axes of rotation of heavy planets are not perpendicular to the
Solar equatorial plane.} with different angles for different Solar-like
systems. These ``orderly'' facts make nebular origin of our Solar system
questionable. To strengthen the doubt further we would like to mention that
for the exoplanets\footnote{E.g. see http://exoplanets.org/} it is not
uncommon to observe planets rotating in the ``wrong'' direction around the
respective stars\footnote{E.g. read \ ``Retrograde motion'' in Wikipedia}. This
trend goes even further to objects such as galaxies\footnote{Disscussed in
section 4.3.}. In spiral galaxies the central bulge typically co-rotates with
the disc. But for the galaxy NGC7331 the bulge and the disc are rotating in
the opposite directions.

In the light of astronomical facts just described Table 1 requires some
extension. For instance, to account for the fact that all planets and regular
satellites are moving in the respective equatorial planes requires us to use
the effects of spin-orbital interactions. Surprisingly, these effects exist in
both Newtonian [$20$] and \ Einsteinian [$83$] gravities\footnote{Read also
note added in proof.}.\ \ At the classical (Newtonian) level the most famous
example of spin-orbital resonance (but of a different kind) is exhibited by
the motion of the Moon whose orbital period coincides with its rotational
period so that it always keeps only one face towards the Earth. Most of the
major natural satellites are locked in analogous 1:1 spin-orbit resonance with
respect to the planets around which they rotate. Mercury represents an
exception since it is locked into 3:2 resonance around the Sun (that is
Mercury completes 3/2 rotation around its axis while making one full rotation
along its orbit) [$20$]. Goldreich [$72$]\ explains such resonances as results
of influence of dissipative (tidal) processes on evolutionary dynamics of
Solar system. The resonance structures observed in the sky are stable
equilibria in the appropriately chosen reference frames [$20$].\ \ Clearly,
the spin-orbital resonances just described\ are not explaining many things.
For instance, while nicely explaining why our Moon is always facing us with
the same side, the same pattern is not observed for Earth rotating around the
Sun with exception of Mercury. Mercury is treated as pointlike object in
canonical general relativity. The spin-orbital interaction [$83$] causing
planets and satellites of heavy planets to lie in the equatorial plane is
different from that causing \ 1:1, etc. resonances. It is analogous to the
NMR-type resonances in atoms and molecules where in the simplest case we are
dealing, say, with the hydrogen atom. \ In it, the proton having spin 1/2 \ is
affected by the magnetic field created by the \ orbital s-electron. In the
atomic case due to symmetry of electron s-orbital this effect is negligible
but nonzero! This effect is known in chemical literature as ``chemical
shift''.\ In the celestial case situation is similar but the effect is expected
to be much stronger since the orbit is planar (not spherical as for the
s-electron in hydrogen atom.\ Hence, the equatorial location of planetary
orbits and regular satellites is likely the result of such spin-orbital
interactions. The equatorial plane in which planets (satellites) move can be
considered as some kind of an orbital (in terminology of atomic physics).\ It
is being filled in accordance with the equivalent of the Pauli
principle:\textsl{\ each orbit can be occupied by no more than one
planet}\footnote{The meteorite belt can be looked upon as some kind of a ring.
We discussed such \textsl{model} rings in Section 3 and in that section the
gravity coming from rings was equated to the gravity coming from planets. We
shall discuss the \textsl{real} rings\ (say, of Saturn, etc.) below, in the
next subsection.}. Once the orbital is filled, other orbitals (planes) will
begin to be filled out. Incidentally, such a requirement automatically
excludes Pluto from the status of a planet. Indeed, on one hand, the T-B-type
law, can easily accommodate Pluto, on another, not only this would contradict
the data summarized in Table 1 and results of previous subsection but also,
and more importantly, \ it would be in contradiction with the astronomical
data for Pluto. According to these data the orbit inclination for Pluto is
17$^{o}$ as compared to the rest of planets whose inclination is within
boundary margins of $\pm$2$^{o}($ except for Mercury for which it is 7$^{o})$.
Some of the orbitals can be empty and \textsl{not all orbits belonging to the
same orbital (a plane) must be filled} (\textsl{as it is also the case in
atomic physics}). This is indeed observed in the sky [$74,77$] and is
consistent with results of \ Table 2. It should be said though that it appears
(according to available data, that not all of the observed satellites are
moving on stable orbits. It appears\ also as if and when the ``inner shell'' is
completely filled, it acts as some kind of an s-type spherical orbital since
\textsl{the} \textsl{orbits of} \textsl{other (irregular) satellites lie
strictly outside the sphere whose diameter is greater or equal to that
corresponding to the last allowed energy level in the first shell}. Such a
situation resembles that for the photon sphere accommodating massless photons
and more distant-massive orbits- discussed in Section 3. \ The location of the
secondary planes \ appears to be quite arbitrary as well as the filling of
their stable orbits. Furthermore, \ without account for spin-orbital
interactions, one can say nothing about the direction of orbital rotation.
Evidently, the ``chemical shift'' created by the motion of regular satellites
lying in the s-shell is such that it should be more energetically advantageous
to rotate in the opposite direction for irregulars. This proposition requires
further study. In addition to planets and satellites on stable orbits there
are many strangers in the Solar system: comets, meteorites, etc. These are
moving not on stable orbits and, as result, should either leave the Solar
system or \ eventually collide with those which move on ``legitimate'' orbits.

\ It is tempting to extend the picture just sketched beyond the scope of our
Solar System. If for a moment we ignore relativistic effects, we can then find
out that our Sun-a typical star in a typical galaxy as explained in section
4.3., is moving along almost circular (Bertrand-type) orbit around our galaxy
center with the period $T=185\cdot10^{6}$ years [$84$]. Our galaxy is also
flat as our Solar system. Hence, again, \ if we believe that stable stellar
motion \ is taking place along geodesics around the galaxy center, then \ we
have to accept that our galaxy is also a quantum object. It would be very
interesting to estimate the number of allowed energy levels (stable orbits)
for our galaxy and to check if the Pauli-like principle works for our and
other galaxies. Some additional thoughts are presented in the next subsection.

\subsection{Role of General Relativity in  the Restricted 3-body Problem and in\\ Dynamics of Planetary Rings}

\ \ \ \ Although literature on the restricted 3-body problem is huge,\ we
would like to discuss this problem from the point of view of \ its connection
with general relativity and quantization of planetary orbits \ along the lines
developed in this paper. We begin with several remarks. First, the existence
of ring systems for \textsl{all} heavy planets is well documented [$74$].
Second, these ring systems are interspersed with satellites of these planets.
Third, both rings and satellites lie in the respective equatorial planes (with
exception of Phoebe's ring) so that satellites move on stable orbits. From
these observations it follows that: \ \ \ \ 

a) \ While each of heavy planets is moving along the geodesics around the Sun,
the respective satellites are moving along the geodesics around \ respective planets.

b) \ The motion of these satellites is almost circular (the condition which
Laplace took into account while studying Jupiter's \ regular satellites).

The restricted 3-body problem can be formulated now as follows. Given that the
rings are made of some kind of small objects whose masses can be
neglected\footnote{This approximation is known as Hill's problem/approximation
in the restricted 3-body problem [$27,46].$} as compared to masses of both
satellite(s) and the respective heavy planet, we can ignore mutual
gravitational \ interaction between these objects (as Laplace did). Under such
conditions we end up with the motion of a given piece of a ring (of zero mass) in
the presence of two bodies of masses $m_{1}$ and $m_{2}$ respectively (the
planet and one of the regular satellites). To simplify matters, it is usually
being assumed that the motion of these two masses takes place on a\ circular
orbit with respect to their center of mass. Complications associated with the
eccentricity of such a motion are discussed in the book by Szebehely [$85$]
and can be taken into account if needed. They will be ignored nevertheless in
this discussion since we shall assume that satellites of heavy planets move on
geodesics so that the center of mass coincides with the position of a heavy
planet thus making our computational scheme compatible with \ methods of
Einstein's relativity. By assuming that ring pieces are massless we also are
making their motion compatible with requirements of general relativity.

Thus far only the motion of \ regular satellites in equatorial planes (of
respective planets) was considered as stable (and, hence, quantizable). The
motion of ring pieces was not accounted for by these stable orbits. The task
now lies in showing that satellites \textsl{lying inside} \textsl{the
respective rings} \textsl{of heavy planets \ are essential for stability of
\ motion of these rings thus making them} \textsl{quantizable}. For the sake
of space, we would like only to provide a sketch of arguments leading to such
a conclusion.

Our task is greatly simplified by the fact that very similar situation exists
for the 3-body system such as Moon, Earth and Sun. \ Dynamics of such a system
was studied thoroughly by Hill whose work played pivotal role in
Poincare$^{\prime}$ studies of celestial mechanics \ [$25$]. Avron and
Simon\textbf{\ [}$86$\textbf{]} adopted Hill's ideas in order to develop
formal quantum mechanical treatment of the Saturn rings. In this work we
follow instead the original Hill's ideas about dynamics of the Earth-Moon-Sun
system first. When these ideas are looked upon from the point of view of
modern mathematics of exactly integrable systems, they enable us to describe
not only the Earth-Moon-Sun system but also the dynamics of rings of heavy
planets. These mathematical methods allow us to find a place for the Hill's
theory within general quantization scheme discussed in previous sections.

To avoid repetitions, we refer our readers to the books of Pars [$50$] and
Chebotarev [$84$] for detailed and clear account of the restricted 3-body
problem and Hill's contributions to Lunar theory. In the nutshell his method
of studying the Lunar problem can be considered as extremely sophisticated
improvement of previously mentioned Laplace method. Unlike Laplace, Hill
realized that both Sun and Earth are surrounded by the rings of
influence\footnote{Related to the so called Roche limit [$74$].}.
The same goes for all heavy planets. Each of these planets and each satellite
of such a planet will have its own domain of influence whose actual width is
controlled by the Jacobi integral of motion.

For the sake of argument, consider the Saturn as an example. It has Pan as its
the innermost satellite. Both the Saturn and the Pan have their \ respective
domains of influence. Naturally, we have to look first at the domain of
influence for the Saturn. Within such a domain let us consider a hypothetical
closed Kepler-like trajectory. \ Stability of such a trajectory is described
by the Hill equation\footnote{In fact, there will be the system of Hill's
equations in \ general [$84$]. This is so since the disturbance of trajectory
is normally decomposed into that which is perpendicular and that \ which is
parallel to  \ Kepler's trajectory at a given point. We shall avoid these
complications though in our work.}. Since such an equation describes a
wavy-type oscillations around the presumably stable trajectory, the parameters
describing such a trajectory are used as an input (perhaps, with subsequent
adjustment) in the Hill equation given by%
\begin{equation}
\frac{d^{2}x}{dt^{2}}+(q_{0}+2q_{1}\cos2t+2q_{2}\cos4t+\cdot\cdot\cdot)x=0.
\tag{77}%
\end{equation}
If at first we would ignore all terms except $q_{0}$, we would \ naively
obtain: $x_{0}(t)=A_{0}cos$($t\sqrt{q_{0}}+\varepsilon).$ This result
describes oscillations around the equilibrium position along the trajectory
with the constant $q_{0}$ carrying information about this trajectory. The
amplitude $A$ is expected to be larger or equal to the average distance
between the pieces of the ring. This naive picture gets very complicated at
once should we use the obtained result as an input into Eq.(77). In this case
the following equation is obtained
\begin{equation}
\frac{d^{2}x}{dt^{2}}+q_{0}x+A_{0}q_{1}\{\cos[t(\sqrt{q_{0}}+2)+\varepsilon
]+\cos[t(\sqrt{q_{0}}-2)-\varepsilon]\}=0 \tag{78}%
\end{equation}
whose solution will enable us to determine $q_{1}$and $A_{1}$ using the
appropriate boundary conditions. Unfortunately, since such a procedure should
be repeated \ infinitely many times, it is obviously impractical. Hill was
able to design a much better method. \ Before discussing Hill's equation from
the perspective of modern mathematics, it is useful \ to recall the very basic
classical facts \ about this equation summarized in the book by Ince [$87$].
For this purpose, we shall assume that solution of Eq.(77) can be presented in
the form
\begin{equation}
x(t)=e^{\alpha t}%
{\textstyle\sum\limits_{r=-\infty}^{\infty}}
b_{r}e^{irt}. \tag{79}%
\end{equation}
Substitution of this result into Eq.(77) leads to the following infinite
system of linear equations%
\begin{equation}
(\alpha+2ri)^{2}b_{r}+%
{\textstyle\sum\limits_{k=-\infty}^{\infty}}
q_{k}b_{r-k}=0,\text{ }r\in\mathbf{Z.} \tag{80}%
\end{equation}
As in finite case, obtaining of the nontrivial solution requires the infinite
determinant $\Delta(\alpha)$ to be equal to zero. \ This problem can be looked
upon from two directions: either all constants $q_{k}$ are assigned and one is
interested in the bounded-type solution for $x(t)$ for $t\rightarrow\infty$
or, one is interested in the relationship between the constants \ made in such
a way that $\alpha=0.$ In the last case it is important to know whether there
is one or more than one of such solutions are available. Although answers can
be found in the book by Magnus and Winkler [$88$], we follow McKean and
Moerbeke [$89$], Trubowitz [$90$] and Moser [$91$] instead. For this purpose,
we need to bring our notations in accord with those used in these references.
Thus, the Hill operator is defined now as $Q(q)=-\frac{d^{2}}{dt^{2}}+q(t)$
with periodic potential $q(t)=q(t+1).$ Eq.(77) can now be rewritten as
\begin{equation}
Q(q)x=\lambda x. \tag{81}%
\end{equation}
Since this is the second order differential equation, it has formally 2
solutions. These solutions depend upon the boundary conditions. For instance,
for \textsl{periodic} solutions such that $x(t)=x(t+2)$ the ``spectrum'' of
\ Eq.(81) is discrete and is given by
\[
-\infty<\lambda_{0}<\lambda_{1}\leq\lambda_{2}<\lambda_{3}\leq\lambda
_{4}<\cdot\cdot\cdot\uparrow+\infty.
\]
We wrote the word spectrum in quotation marks because of the following.
Eq.(81) does have a normalizable solution only if $\lambda$ belongs to the
(pre assigned) intervals $(\lambda_{0},\lambda_{1}),\break (\lambda_{2},\lambda
_{3}),...,(\lambda_{2i},\lambda_{2i+1}),...$ \ \ In such a case the
eigenfunctions $x_{i}$ are normalizable in the usual sense of quantum
mechanics and form the orthogonal set. Periodic solutions make sense only for
vertical displacement from the reference trajectory. For the horizontal
displacement the boundary condition should be chosen as $x(0)=x(1)=0.$ For
such chosen boundary condition the discrete spectrum also exists but it lies
exactly in the gaps between the intervals just described, i.e. $\lambda
_{1}\leq\mu_{1}\leq\lambda_{2}<\lambda_{3}\leq\mu_{2}\leq\lambda_{4}%
<\cdot\cdot\cdot.$ For such a spectrum there is also set of normalized
mutually orthogonal eigenfunctions. \textsl{Thus in both cases quantum
mechanical} \textsl{description is assured}. One can do much more however. In
particular, Trubowitz [$90$] designed an explicit procedure for recovering the
potential $q(t)$ from the $\mu-$spectrum supplemented by the information about
normalization constants.

\ \ It is quite remarkable that the Hill equation can be interpreted in terms
of the auxiliary dynamical (Neumann) problem. Such an interpretation is very
helpful for the purposes of this work since it allows to include the quantum
mechanical treatment of Hill's equation into general framework developed in
this paper.

Before describing such a connection, we would like to add few details to
results of previous subsection. First, as in the planetary case, the number of
pre assigned intervals is always finite. This means that, beginning with some
pre assigned $\hat{\imath}$, we would be left with $\lambda_{2i}%
=\lambda_{2i+1}\forall i>\hat{\imath}.$ These \textsl{double} eigenvalues do
not have independent physical significance since they can be determined by the
set of \textsl{single} eigenvalues (for which $\lambda_{2i}\neq\lambda
_{2i+1})$ as demonstrated by Hochstadt [$92$]. Because of this, the potentials
$q(t)$ in Hill's equation are called the \textsl{finite gap}
potentials\footnote{Since there is only finite number of gaps [$\lambda
_{1},\lambda_{2}],$[$\lambda_{3},\lambda_{4}],...$where the spectrum is
forbidden.}. Hence, physically, it is sufficient to discuss only potentials
which possess finite single spectrum. The auxiliary $\mu-$spectrum is then
determined by the gaps of the single spectrum as explained above.

With this information in our hands, we are ready to discuss the exactly
solvable Neumann dynamical problem. It is the problem about dynamics of a
particle moving on $n-$dimensional sphere $<\mathbf{\xi},\mathbf{\xi}%
>\equiv\xi_{1}^{2}$ +$\cdot\cdot\cdot+\xi_{n}^{2}=1$ under the influence of a
quadratic potential $\phi(\mathbf{\xi})=<\mathbf{\xi},\mathbf{A\xi}>.$
Equations of motion describing the motion on $n-$ sphere are given by
\begin{equation}
\mathbf{\ddot{\xi}}=-\mathbf{A\xi}+u(\mathbf{\xi})\mathbf{\xi}\text{ \ with
}u(\mathbf{\xi})=\phi(\mathbf{\xi})-<\mathbf{\dot{\xi}},\mathbf{\dot{\xi}}>.
\tag{82}%
\end{equation}
Without loss of generality, we assume that the matrix $\mathbf{A}$ is already
in the diagonal form: $\mathbf{A}:=diag(\alpha_{1},...,\alpha_{n}).$ With such
an assumption we can equivalently rewrite Eq.(68) in the following suggestive
form%
\begin{equation}
\left(  -\frac{d^{2}}{dt^{2}}+u(\mathbf{\xi}(t))\right)  \xi_{k}=\alpha_{k}%
\xi_{k}\text{ ; \ }k=1,...,n. \tag{83}%
\end{equation}
Thus, in the case if we can prove that $u(\mathbf{\xi}(t))$ in \ Eq.(83) is
the same as $q(t)$ in Eq.(81), the connection between the Hill and Neumann's
problems will be established. The proof is given in Appendix C. It is
different from that given in the lectures by Moser [$91$] since it is more
direct and much shorter.

This proof brought us unexpected connection with hydrodynamics through \ the
static version of the Korteweg-de Vries equation. Attempts to describe the
Saturnian rings using equations of hydrodynamic are described in the recent
monograph by Esposito [$93$]. This time, however, we can accomplish more using
\ already obtained information.

Following Kirillov [$94$], we introduce the commutator for the fields
(operators) $\xi$ and $\eta$ as follows: $[\xi,\eta]=\xi\partial\eta
-\eta\partial\xi.$ Using the KdV equation (C.10), let us consider 3 of its
independent solutions: $\xi_{0},\xi_{-1}$ and $\xi_{1}.$ All these solutions
can be obtained from general result: $\xi_{k}=t^{k+1}+O(t^{2}),$ valid near
zero. Consider now the commutator $[\xi_{0},\xi_{1}].$ Straightforwardly, we
obtain, $[\xi_{0},\xi_{1}]=\xi_{1}$. Analogously, we obtain, $[\xi_{0}%
,\xi_{-1}]=-\xi_{-1}$ and, finally, $[\xi_{1},\xi_{-1}]=-2\xi_{0}.$ According
to Kirillov, such a Lie algebra is isomorphic to that for the group $SL(2,R)$
which is the center for the Virasoro algebra\footnote{Since connections
between the KdV and the Virasoro algebra are well documented [$95$], it is
possible in principle to reinterpret fine structure of the Saturn's rings
string-theoretically.}. Vilenkin [$96$] demonstrated that the group $SL(2,R)$
is isomorphic to $SU(1,1)$. Indeed, by means of transformation: \textit{w}%
\textsl{=}$\dfrac{\mathit{z}-i}{\mathit{z}+i},$ it is possible to transform
the upper half plane (on which $SL(2,R)$ acts) into the interior of unit
circle on which $SU(1,1)$ acts. Since, according to Appendix A, the group
$SU(1,1)$ is the connected component of $SO(2,1)$, the anticipated connection
with the $SO(2,1)$ group (discussed already) is established.

In Appendix C we noticed connections between the Picard-Fuchs, Hill and
Neumann-type equations. In a recent paper by Veselov et al [97] such a
connection was developed much further resulting in the Knizhnik-Zamolodchikov-
type equations for the Neumann-type dynamical systems. We refer our readers to
the original literature, especially to the well written lecture notes by Moser
[$91$]. These notes as well and his notes in collaboration with Zehnder [$32$]
provide an excellent background for study the whole circle of ideas discussed
in this subsection.

\subsection{Using Contact Geometry, Topology and Quantum   Mechanics for\\ Calculation of Rotation Curves for Typical Stars in Spiral Galaxies}

To make a progress with calculation of the rotation curve, we need to return
to the original source.That is to the paper by Perlick [$69$]. Using equations
(57)-(60) and following protocol of classical mechanics we substitute
equations (58)-(60) into (57). The following result is then obtained%
\begin{equation}
\dot{r}^{2}=e^{-\lambda(r)}(2E-\frac{M^{2}}{r^{2}}+e^{-\nu(r)}) \tag{84}.%
\end{equation}
By comparing this result with standard classical mechanics results for the
particle motion in central field [$65$], we should interpret the factor
$-e^{-\nu(r)}$ as potential $U(r)$. Detailed analysis of this mechanical
problem done by Perlic \  resulted in the following conclusions.

1. To study particle dynamics it is sufficient to study point-like
nonrelativistic particle moving on a Riemannian 3-manifold with metric
$e^{\lambda(r)}dr^{2}+r^{2}(d\theta^{2}+\sin^{2}\theta d\varphi^{2})$ in the
presence of a potential $U(r)=-e^{-\nu(r)}.$ It is clear, that this metric is
reducible to that for the 3- sphere $S^{3}$[$69$].

2. In view of spherical symmetry, the factor $e^{\lambda(r)}$ can be chosen,
for instance, as $(1-\frac{r^{2}}{4})^{-1}$, which is appropriate for the
standard metric on $SO(3)$. With such a choice, the potential $U(r)$ will
\ look either as $U_{1}(r)=-G+2r^{2}(4-r^{2})^{-1}$ or as $U_{2}%
(r)=-G-\frac{1}{r}\sqrt{1-4r^{2}}$ with $G$ being some nonnegative constant.
\ Evidently, $U_{1}(r)$ is \ preferable for large r's while $U_{2}(r)$
should be used for small r's. \ In the first case we obtain: $U_{1}%
(r\rightarrow\infty)=-G-2$, while in the second, $U_{2}(r)=-G-$ $\frac{1}%
{r}+2r.$ In the limit $r\rightarrow\infty$ the potential becomes a constant
and the dynamical problem is reduced to that of finding closed geodesics on
$SO(3).$

3. \ Since the rotational group $SO(3)=SU(2)/[I,-I]$, [$98$], in the limit
$r\rightarrow\infty$ it is permissible to consider the dynamics on $SU(2)$.

\bigskip

Fortunately, the dynamics on $SU(2)$ was discussed in detail in our book
[$33$], section 6.5. Referring to the book for details, here we provide only
the condensed summary. First of all we notice that $S^{3}$ can be represented
as Hopf fibration : $S^{2}=S^{3}/S^{1}$ this can be easily understood as
follows. Since $SU(2)=S^{3}$ and since $S^{3}$ can be respresented as Bloch
sphere $\left\vert z_{1}\right\vert ^{2}$+$\left\vert z_{2}\right\vert ^{2}$=1
we observe that $S^{2}=\frac{z_{2}}{z_{1}}.$ This is so because $z_{1}%
=a_{1}+ib_{1}=r_{1}e^{i\phi_{1}}$ and, accordingly, $z_{2}=a_{2}+ib_{2}%
=r_{2}e^{i\phi_{2}}.$ Therefore, $S^{2}=\frac{z_{2}}{z_{1}}=\frac{r_{2}}%
{r_{1}}e^{i(\phi_{2}-\phi_{1})}$ . That is $S^{2}$ is the complex z-plane +
point at infinity.This can be seen using the stereographic projection. Thus
dynamics on $S^{3}$ can be reduced to that in the plane in the presence of the
area constraint. The Lagrangian for such a case can be written as
\begin{equation}
\mathcal{L(}x_{1},x_{2},t;\dot{x}_{1},\dot{x}_{2},\dot{t})=\frac{1}{2}(\dot
{x}_{1}^{2}+\dot{x}_{1}^{2})+\lambda(\dot{t}-2x_{2}\dot{x}_{1}+2x_{1}\dot
{x}_{2}). \tag{85}%
\end{equation}
Here the first term represents kinetic energy while the second the area
constraint. $\lambda$ is the Lagrange multiplier. \ Notice that the above
Lagrangian is written in the extended configuratioin space (by doing so we are
entering the domain of general relativity (read section 4.3)).Consider now a
curve \textbf{c(}$\tau)$ in configurational space \textbf{c(}$\tau)$%
=\{x$_{1}(\tau)$,x$_{2}(\tau)$,t($\tau)$\}. The velocity of this curve is
given by%
\begin{equation}
\mathbf{\dot{c}(}\tau)=\dot{x}_{1}\partial_{x_{1}}+\dot{x}_{2}\partial_{x_{2}%
}+\dot{t}\partial_{t}=\dot{x}_{1}X_{1}+\dot{x}_{2}X_{2}+(\dot{t}-2x_{2}\dot
{x}_{1}+2x_{1}\dot{x}_{2})\partial_{t} \tag{86a}%
\end{equation}
The curve \textbf{c(}$\tau)$ is \textsl{horizontal} if
\begin{equation}
\dot{t}=2x_{2}\dot{x}_{1}-2x_{1}\dot{x}_{2} \tag{86b}%
\end{equation}
Furthermore, $X_{1}=\partial_{x_{1}}+2x_{2}\partial_{t},X_{2}=\partial_{x_{1}%
}-2x_{1}\partial_{t}$ and $[X_{1},X_{2}]=-4\partial_{t}.$ Working in the
Hilbert space of wave functions of the type $\psi=(x_{1},x_{2})e^{-\frac
{it}{4}}$ the commutator just defined becomes ($\hbar=1):$
\begin{equation}
\lbrack X_{1},X_{2}]=i \tag{86c}%
\end{equation}
that is easily recognizable as Heisenberg's fundamental commutation relation
of quantum mechanics. Furthermore, it is direct consequence of the
horizontality condition (86b)! To demonstrate this, we rewrite the equation
(86b) as
\begin{equation}
-dt+2(x_{2}dx_{1}-2x_{1}dx_{2})=0. \tag{87a}%
\end{equation}
Define 1-form $\alpha$ via
\begin{equation}
\alpha=-\frac{1}{4}dt+\frac{1}{2}(x_{2}dx_{1}-x_{1}dx_{2}), \tag{87b}%
\end{equation}
then the horizontality condition (87a) can be equivalently rewritten as
\begin{equation}
\alpha(X_{i})=0,i=1,2. \tag{87c}%
\end{equation}
Thus, the ghorizontality condition (87a) is completely equivalent to
Heisenberg's commutation relation. Furthermore, going back to equation (49) we
now can easily recognize in it the form analogous to $\alpha,$defined by (87b)
. Equivalently, the energy constraint (53) can be rewritten as the
horizontality condition (87a). Indeed,
\begin{equation}%
{\displaystyle\sum\limits_{i,k}}
g_{ik}(q)\frac{dq^{i}}{d\tau}dq^{k}-(E-V(q)d\tau=0 \tag{88}%
\end{equation}
From here we conclude that the motion of \ the point constrained to the energy
surface in configuration space is nesessarily quantized. In the present case
the problem of star motion on the circular orbit around galaxy center is
exactly equivalent to the Landau-type problem about motion of nonrelativistic
electron in constant magnetic field. \ In magnetic field the kinetic energy is
a constant of motion as well as the magnitude of the velocity. However, since
the motion is quantized, the radii $r_{n}^{2}$ of the Landau orbits grow as
\begin{equation}
r_{n}^{2}=const\text{ n} \tag{89}%
\end{equation}
Equivalently, this can be reinterpreted in terms of the flux quantization. The
$const$ factor includes the crossover scale $r_{0}$ at which \ the rotation
curve crossovers from the regime $\left\vert \mathbf{v}\right\vert =Ar,r<$
$r_{0}$ ($A$ is some constant) to the regime $\left\vert \mathbf{v}\right\vert
=const$, $r\geq r_{0}.$ Analogy with magnetism allows us to restore the exact
values of all constant involved. This task is left for future work.

\section{Concluding Remarks}

\ Although Einstein was not happy with the existing formulation of quantum
mechanics, the results obtained in this work demonstrate harmonious
coexistence of general relativity and quantum mechanics.\ It should be noted
though that such a harmony had been achieved at the expense of partial
sacrificing of the correspondence principle. This principle is not fully
working. Even for such well studied system as Hydrogen atom as discussed in
subsection 4.2. there are some \ logical problems. This fact should not be
considered as too worrisome as it was to Einstein. Indeed, as Heisenberg
correctly pointed out: all what we know about microscopic system is its
spectrum (in the very best of cases). \ We argued that at the scales of Solar
system the existing intrinsic uncertinty in any measurement blures away
differences between classical and quantum formalisms. This conclusion can be
reached independently based on considerations of contact geometry and topology
discussed in section 5.6. \ Based on results of this work, contact geometry,
also known as sub-Riemannian geometry [$33$], can be looked upoon as yet
another branch of general relativity.

\bigskip

\textbf{Note added in proof.} When this paper was completed, the following
papers came to our attention. 1. In the paper arXiv:1104.0548 ``On the first
determination of Mercury's perihelion advance'' \ Diana Rodica Constantin
\ recalculated Leverrier's results with surprising outcome. Correction to the
perihelion shift of Mercury was found as 42\textquotedblright.8/century as
compared to 42\textquotedblright.98/century obtained in general relativity, 2.
In the paper arXiv:1008.1811 ``General Relativity Problem of Mercury's
Perihelion Advance Revisited'' by Anatoli Vankov Le Verrier-type calculations
of the type presented in our section 3.1 are discussed with more detail and
more historical background, 3. In the paper ``The geometry of photon surfaces''
by C-M. Claudel, K. S. Virbhadra and G. F. R. Ellis, J.Math.Phys.\textbf{42},
818 (2001), the concept of a photon sphere in Schwarzschild space--time
discussed in Ref.[$10$] of the main text is generalized to a definition of a
photon surface in an arbitrary space--time. \bigskip The book [$99$] contains
a very detailed list of people, begining with Leverrier, who calculated the
perihelion shift of Mercury. Surprisingly, this book does not contain works by
O.Heaviside who had also calculated the perihelion shift of Mercury in 1893.
His calculations were reconsidered and extended by Jefimenko. His approach to
general relativity based on Heaviside ideas is presented in the monograph
[$100$] which also includes the original papers by Heaviside\footnote{The
author would like to thank the anonimous referee for these references.}. Last
but not the least, the idea of forceless classical mechanics is attributed to
H.Hertz $[64$], pages 130-132. The force-free geodesic motion in classical
mechanics was actually championed by Maupertuis already in 18th century
\ between 1741 and
1746\footnote{http://en.wikipedia.org/wiki/Pierre\_Louis\_Maupertuis} . \ 

\bigskip

\ \ \ \ \ \ \ \ \ {\Large Appendix A: Some Quantum Mechanical Problems}

\ \ \ \ \ \ \ \ \ {\Large Associated With the Lie Algebra of SO(2,1)
Group\bigskip\bigskip}

Following Wybourne [$56$] consider the second order differential equation of
the type%
\begin{equation}
\frac{d^{2}Y}{dx^{2}}+V(x)Y(x)=0 \tag{A.1}%
\end{equation}
where $V(x)=a/x^{2}+bx^{2}+c.$ Consider as well the Lie algebra of the
noncompact group SO(2,1) or, better, its connected component SU(1,1). It is
given by the following commutation relations%
\begin{equation}
\lbrack X_{1},X_{2}]=-iX_{3};\text{ }[X_{2},X_{3}]=iX_{1};\text{ }[X_{3}%
,X_{1}]=iX_{2} \tag{A.2}%
\end{equation}
We shall seek the realization of this Lie algebra in terms of the following
generators%
\begin{equation}
X_{1}:=\frac{d^{2}}{dx^{2}}+a_{1}(x);\text{ \ }X_{2}:=i[k(x)\frac{d}{dx}%
+a_{2}(x)];\text{ \ }X_{3}:=\frac{d^{2}}{dx^{2}}+a_{3}(x). \tag{A.3}%
\end{equation}
The unknown functions $a_{1}(x),a_{2}(x),a_{3}(x)$ and $k(x)$ are determined
upon substitution of (A.3) into (A.2). After some calculations, the following
result is obtained%
\begin{equation}
X_{1}:=\frac{d^{2}}{dx^{2}}+\frac{a}{x^{2}}+\frac{x^{2}}{16};\text{ }%
X_{2}:=\frac{-i}{2}[x\frac{d}{dx}+\frac{1}{2}];\text{ }X_{3}:=\frac{d^{2}%
}{dx^{2}}+\frac{a}{x^{2}}-\frac{x^{2}}{16}. \tag{A.4}%
\end{equation}
In view of this, (A.1) can be rewritten as follows
\begin{equation}
\lbrack(\frac{1}{2}+8b)X_{1}+(\frac{1}{2}-8b)X_{3}+c]Y(x)=0. \tag{A.5}%
\end{equation}
This expression can be further simplified by the unitary transformation$UX_{1}%
U^{-1}=X_{1}\cosh\theta+X_{3}\sinh\theta;$ $UX_{3}U^{-1}=X_{1}\sinh
\theta+X_{3}\cosh\theta$ with $U=exp(-i\theta X_{2}).$ By choosing
$\tanh\theta=-(1/2+8b)/(1/2-8b)$ (A.5) is reduced to
\begin{equation}
X_{3}\tilde{Y}(x)=\frac{c}{4\sqrt{-b}}\tilde{Y}(x), \tag{A.6}%
\end{equation}
where the eigenfunction $\tilde{Y}(x)=UY(x)$ is an eigenfunction of both
$X_{3}$ and the Casimir operator \textbf{X}$^{2}=X_{3}^{2}-X_{2}^{2}-X_{1}%
^{2}$ so that by analogy with the Lie algebra of the angular momentum we
obtain,
\begin{align}
\mathbf{X}^{2}\tilde{Y}_{jn}(x)  &  =J(J+1)\tilde{Y}_{Jn}(x)\text{
\ \ and}\tag{A.7a}\\
X_{3}\tilde{Y}_{Jn}(x)  &  =\frac{c}{4\sqrt{-b}}\tilde{Y}_{Jn}(x)\equiv
(-J+n)\tilde{Y}_{Jn}(x)\text{; }\ n=0,1,2,...\text{.} \tag{A.7b}%
\end{align}
It can be shown that $J(J+1)=-a/4-3/16$. From here we obtain : $J=-\frac{1}%
{2}(1\pm\sqrt{\frac{1}{4}-a});$ $\frac{1}{4}-a\geq0.$ In the case of discrete
spectrum one should choose the plus sign in the expression for $J$. Using this
result in (A.7) we obtain the following result of major importance%

\begin{equation}
4n+2+\sqrt{1-4a}=\frac{c}{\sqrt{-b}}. \tag{A.8}%
\end{equation}
Indeed, consider the planar Kepler problem. In this case, in view of (3.5),
the radial Schr\"{o}dinger equation can be written in the following symbolic
form%
\begin{equation}
\left[  \frac{d^{2}}{dr^{2}}+\frac{1}{r}\frac{d}{dr}+\frac{\mathit{\upsilon}%
}{r}+\frac{u}{r^{2}}+g\right]  R(r)=0 \tag{A.9}%
\end{equation}
By writing $r=x^{2}$ and $R(r)=x^{-\frac{1}{2}}\mathcal{R}(x)$ This equation
is reduced to the canonical form given by (A.1), e.g. to%
\begin{equation}
(\frac{d^{2}}{dx^{2}}+\frac{4u+1/4}{x^{2}}+4gx^{2}+4\upsilon)\mathcal{R}(x)=0
\tag{A.10}%
\end{equation}
so that the rest of arguments go through. Analogously, in the case of
Morse-type potential we have the following Schrodinger-type equation
initially:%
\begin{equation}
\left[  \frac{d^{2}}{dz^{2}}+pe^{2\alpha z}+qe^{\alpha z}+k\right]  R(z)=0
\tag{A.11}%
\end{equation}
By choosing $z=lnx^{2}$ and $R(z)=x^{-\frac{1}{2}}\mathcal{R}(x)$ (A11) is
reduced to the canonical form%
\begin{equation}
(\frac{d^{2}}{dx^{2}}+\frac{16k+\alpha^{2}}{4\alpha^{2}x^{2}}+\frac{4p}%
{\alpha^{2}}x^{2}+\frac{4q}{\alpha^{2}})\mathcal{R}(x)=0. \tag{A.12}%
\end{equation}
By analogous manipulations one can reduce to the canonical form the radial
equation for Hydrogen atom and for 3-dimensional harmonic oscillator.

\bigskip\bigskip

{\Large Appendix B: Numerical Data Used for Claculations of}

{\Large n}$_{theory}^{\ast}${\Large (Supplement to Table 2)}

\bigskip

\ 

1 au=149.598$\cdot10^{6}km$

\smallskip

Masses (in kg): Sun 1.988$\cdot10^{30},$ Jupiter 1.8986$\cdot10^{27}$, Saturn
5.6846$\cdot10^{26},$

Uranus 8.6832$\cdot10^{25},$ Neptune 10.243$\cdot10^{25}.$

\smallskip

q$_{j}:$ Jupiter 0.955$\cdot10^{-3},$ Saturn 2.86$\cdot10^{-4},$ Uranus
4.37$\cdot10^{-5}$, Neptune 5.15$\cdot10^{-5}.$

\smallskip

$\left(  r_{j}\right)  _{1}(km):$ Jupiter 127.69$\cdot10^{3},$ Saturn
133.58$\cdot10^{3},$ Uranus 49.77$\cdot10^{3},$

Neptune 48.23$\cdot10^{3}.$

\smallskip

ln$\left(  \dfrac{\gamma M}{2r_{1}}\right)  $ : Earth 4.0062, Jupiter 3.095,
Saturn 1.844, Uranus 0.9513,

Neptune 1.15.

\medskip\bigskip

{\Large Appendix C: Connections Between the Hill and }

{\Large Neumann's Dynamical Problems}

\bigskip

Following our paper [$101$], let us consider the Fuchsian-type equation given
by
\begin{equation}
y^{^{\prime\prime}}+\frac{1}{2}\phi y=0, \tag{C.1}%
\end{equation}
where the potential $\phi$ is determined by the equation $\phi=[f]$ with
$f=y_{1}/y_{2}$ and $y_{1},y_{2}$ \ being two independent solutions of (C.1)
normalized by the requirement $y_{1}^{^{\prime}}y_{2}$ -$y_{2}^{\prime}%
$\ $y_{1}=1.$The symbol$\ [f]$ denotes the Schwarzian derivative of $f$. Such
a derivative is defined as follows
\begin{equation}
\lbrack f]=\frac{f^{\prime}f^{\prime\prime\prime}-\frac{3}{2}\left(
f^{\prime\prime}\right)  ^{2}}{\left(  f^{\prime}\right)  ^{2}}. \tag{C.2}%
\end{equation}
Consider (C.1) on the circle $S^{1}$ and consider some map of the circle given
by $F(t+1)=F(t)+1.$ Let $t=F(\xi)$ so that $y(t)=Y(\xi)\sqrt{F^{\prime}(\xi)}$
leaves (C.1) form -invariant, i.e. in the form $Y^{\prime\prime}+\frac{1}%
{2}\Phi Y=0$ with potential $\Phi$ being defined now as $\Phi(\xi)=\phi
(F(\xi))[F^{\prime}(\xi)]^{2}+[F(\xi)].$ Consider next the infinitesimal
transformation $F(\xi)=\xi+\delta\varphi(\xi)$ with $\delta$ being some small
parameter and $\varphi(\xi)$ being some function to be determined. Then,
$\Phi(\xi+\delta\varphi(\xi))=\phi(\xi)+\delta(\hat{T}\varphi)(\xi
)+O(\delta^{2}).$ Here $(\hat{T}\varphi)(\xi)=\phi(\xi)\varphi^{\prime}%
(\xi)+\frac{1}{2}\varphi^{\prime\prime\prime}(\xi)+2\phi^{\prime}(\xi
)\varphi(\xi).$ Next, we assume that the parameter $\delta$ plays the same
role as time. Then, we obtain
\begin{equation}
\lim_{t\rightarrow0}\frac{\Phi-\phi}{t}=\frac{\partial\phi}{\partial t}%
=\frac{1}{2}\varphi^{\prime\prime\prime}(\xi)+\phi(\xi)\varphi^{\prime}%
(\xi)+2\phi^{\prime}(\xi)\varphi(\xi) \tag{C.3}%
\end{equation}
Since thus far the perturbing function $\varphi(\xi)$ was left undetermined,
we can choose it now as $\varphi(\xi)=\phi(\xi).$ Then, we obtain the Korteweg
-de Vriez \ (KdV) equation%
\begin{equation}
\frac{\partial\phi}{\partial t}=\frac{1}{2}\phi^{\prime\prime\prime}%
(\xi)+3\phi(\xi)\phi^{\prime}(\xi) \tag{C.4}%
\end{equation}
determining the potential $\phi(\xi).$ For reasons which are explained in the
text, it is sufficient to consider only the static case of KdV, i.e.%
\begin{equation}
\phi^{\prime\prime\prime}(\xi)+6\phi(\xi)\phi^{\prime}(\xi)=0. \tag{C.5}%
\end{equation}
We shall use this result as a reference for our main task of connecting the
Hill and  Neumann's problems. Using \ Eq.(68) we write%
\begin{equation}
u(\xi)=\phi(\xi)-<\dot{\xi},\dot{\xi}>. \tag{C.6}%
\end{equation}
Consider an auxiliary functional $\varphi(\xi)=<\xi,A^{-1}\xi>.$ Suppose that
$\varphi(\xi)=u(\xi).$ Then,
\begin{equation}
\frac{du}{dt}=2<\dot{\xi},A\xi>-2<\ddot{\xi},\dot{\xi}>. \tag{C.7}%
\end{equation}
But $<\ddot{\xi},\dot{\xi}>=0$ because of the normalization constraint
$<\xi,\xi>=1.$ Hence, $\dfrac{du}{dt}=2<\dot{\xi},A\xi>.$ Consider as well
$\dfrac{d\varphi}{dt}.$ By using Eq.(68) it is straightforward to show that
$\dfrac{d\varphi}{dt}=2<\dot{\xi},A^{-1}\xi>.$ Because by assumption
$\varphi(\xi)=u(\xi),$ we have to demand that $<\dot{\xi},A^{-1}\xi>=<\dot
{\xi},A\xi>$ as well. If this is the case, consider
\begin{equation}
\dfrac{d^{2}u}{dt^{2}}=2<\ddot{\xi},A^{-1}\xi>+2<\dot{\xi},A^{-1}\dot{\xi}>.
\tag{C.8}%
\end{equation}
Using Eq.(68) once again we obtain,%
\begin{equation}
\dfrac{d^{2}u}{dt^{2}}=-2+2u\varphi+2<\dot{\xi},A^{-1}\dot{\xi}>. \tag{C.9}%
\end{equation}
Finally, consider as well $\dfrac{d^{3}u}{dt^{3}}.$ Using (C.9) as well as
Eq.(68) and (C.7) we obtain,%
\begin{equation}
\dfrac{d^{3}u}{dt^{3}}=2\frac{du}{dt}\varphi+4u\frac{du}{dt}=6u\frac{du}{dt}.
\tag{C.10}%
\end{equation}
By noticing that in (C.5) we can always make a rescaling $\phi(\xi
)\rightarrow\lambda\phi(\xi),$ we can always choose $\lambda=-1$ so that (C.5)
and (C.10) coincide. This result establishes correspondence between the
Neumann and Hill-type problems.

\end{document}